\documentclass[a4paper,11pt]{article}
\pdfoutput=1 

\usepackage{jheppub} 
                     
\usepackage{graphicx} 
\usepackage{epsfig}
\usepackage{epsf}
\usepackage{epstopdf}
\usepackage{hyperref}
\usepackage{cancel}

\usepackage{amsmath}  
\usepackage{pdfsync}
\usepackage{slashed}

\newcommand{\bea}{\begin{eqnarray}} 
\newcommand{\eea}{\end{eqnarray}} 
\def\eqa{&=&}

\def\ep{\epsilon}

\def\ra{\rangle}

\newcommand{\jhat}[1]{\hspace{0.3em}\widehat{\hspace{-0.4em}#1\hspace{-0.4em}}\hspace{0.4em}}

\newcommand{\symb}{\textrm{symb}}
\newcommand{\symbi}{\symb^{-1}}

\newcommand{\ps}{\slashed{p}}

\newcommand{\ks}{\slashed{k}}

\newcommand{\dps}{\displaystyle}

\usepackage[T1]{fontenc} 

\def\targ{(\tau,\tau')}

\def\argS{(k_1,\varepsilon_1;\ldots ; k_S,\varepsilon_S)}

\def\intT{\int_0^\infty dT\, {\rm e}^{-m^2T}}

\def\eTq{{\rm e}^{-\frac{1}{4}\int_0^Td\tau\,\dot{q}^2}}

\def\t #1{\tau_{#1}}
\def\non{\nonumber\\}

\def\e{\,{\rm e}}

\def\veps#1{\varepsilon_{#1}}

\def\kk#1#2{k_{#1}\cdot k_{#2}}

\def\lin{\Big\vert_{\veps}}
\def\multilin{\Big\vert_{\veps_1\veps_2\cdots \veps_N}}

\def\half{{1\over 2}}

\def\fourth{{1\over4}}

\def\Eins{{\mathchoice {\rm 1\mskip-4mu l} {\rm 1\mskip-4mu l}
{\rm 1\mskip-4.5mu l} {\rm 1\mskip-5mu l}}}
\def\Z{{\mathchoice {\hbox{$\sf\textstyle Z\kern-0.4em Z$}}
{\hbox{$\sf\textstyle Z\kern-0.4em Z$}}
{\hbox{$\sf\scriptstyle Z\kern-0.3em Z$}}
{\hbox{$\sf\scriptscriptstyle Z\kern-0.2em Z$}}}}
\def\abs#1{\left| #1\right|}

        \def\slash#1{#1\!\!\!\raise.15ex\hbox {/}}
\newcommand{\slD}{\,\raise.15ex\hbox{$/$}\kern-.27em\hbox{$\!\!\!D$}}
\newcommand{\slpartial}{\raise.15ex\hbox{$/$}\kern-.57em\hbox{$\partial$}}

\def\ddel{{}^\bullet\! \Delta}
\def\deld{\Delta^{\hskip -.5mm \bullet}}

\def\ddeld{{}^{\bullet}\! \Delta^{\hskip -.5mm \bullet}}

\def\epsk#1#2{\varepsilon_{#1}\cdot k_{#2}}
\def\epseps#1#2{\varepsilon_{#1}\cdot\varepsilon_{#2}}

\def\no{\noindent}

\def\kinq{{1\over 4}\dot q^2}
\def\kinb{{1\over 4}\dot x^2}

\def\der#1#2{{d #1\over d#2}}
\def\partder#1#2{{\partial #1\over\partial #2}}

\def\be{\begin{equation}}
\def\ee{\end{equation}\noindent}
\def\bear{\begin{eqnarray}}
\def\ear{\end{eqnarray}\noindent}
\def\bec{\blue\begin{equation}}
\def\eec{\end{equation}\black\noindent}
\def\bearc{\blue\begin{eqnarray}}
\def\earc{\end{eqnarray}\black\noindent}
\def\benn{\begin{enumerate}}
\def\enn{\end{enumerate}}
\def\ee{&=&}

\def\ab{{\alpha\beta}}

\def\mn{{\mu\nu}}

\def\tr{{\rm tr}\,}

\def\e{\,{\rm e}}

\def\b0{{\bf 0}}

\def\4piTD{{(4\pi T)}^{-{D\over 2}}}
\def\4piT4{{(4\pi T)}^{-2}}

\def\argS{(k_1,\varepsilon_1;\ldots ; k_S,\varepsilon_S)}

\newcommand{\s}{\slashed}
\def\veps{\varepsilon}

\newcommand{\epsilonilon}{\varepsilon}
\def\veps{\epsilonilon}
\newcommand{\bone}{1\!\!1}

\title{\boldmath Worldline master formulas for the dressed electron propagator, part 1: Off-shell amplitudes}

\author[a]{N. Ahmadiniaz,}
\author[f]{V. M. Banda Guzm\'an,}
\author[c,d]{F. Bastianelli,}
\author[e,d]{O. Corradini,}
\author[f]{J.P. Edwards}
\author[f]{and C. Schubert}

\affiliation[a]{ Helmholtz-Zentrum Dresden-Rossendorf, Bautzner Landstra\ss e 400, 01328 Dresden, Germany}
\affiliation[c]{Dipartimento di Fisica ed Astronomia, Universit\`a di Bologna, Via Irnerio 46, I-40126 Bologna, Italy}
\affiliation[d]{INFN, Sezione di Bologna, Via Irnerio 46, I-40126 Bologna, Italy}
\affiliation[e]{Dipartimento di Scienze Fisiche, Informatiche e Matematiche,
 Universit\`a degli Studi di Modena e Reggio Emilia, Via Campi 213/A, I-41125 Modena, Italy}
\affiliation[f]{Instituto de F\'isica y Matem\'aticas
Universidad Michoacana de San Nicol\'as de Hidalgo
Edificio C-3, Apdo. Postal 2-82
C.P. 58040, Morelia, Michoac\'an, M\'exico}
\emailAdd{n.ahmadiniaz@hzdr.de}
\emailAdd{victor.banda@umich.mx}
\emailAdd{bastianelli@bo.infn.it} 
\emailAdd{olindo.corradini@unimore.it}
\emailAdd{jedwards@ifm.umich.mx}
\emailAdd{schubert@ifm.umich.mx}

\abstract{
In the first-quantised worldline approach to quantum field theory, a long-standing problem has been to 
extend this formalism to amplitudes involving open fermion lines while maintaining the efficiency of the well-tested
closed-loop case. In the present series of papers, we develop a suitable formalism for the case of quantum electrodynamics 
in vacuum (part one and two) and in a constant external electromagnetic field (part three), 
based on second-order fermions and the symbol map.
We derive this formalism from standard field theory, but also give an alternative derivation intrinsic to the worldline theory.
In this first part, we use it to obtain a Bern-Kosower type master formula for the fermion propagator, dressed with $N$ photons, in terms
of the ``$N$-photon kernel,'' where off-shell this kernel appears also in ``subleading'' terms involving only $N-1$ of the $N$ photons.
Although the parameter integrals generated by the master formula are equivalent to the usual
Feynman diagrams, they are quite different since the 
use of the inverse symbol map avoids the appearance of long products of Dirac matrices. 
As a test we use the $N=2$ case for a recalculation of the one-loop fermion self energy, in $D$ dimensions and arbitrary covariant
gauge, reproducing the known result. We find that significant simplification can be achieved in this calculation by choosing an unusual momentum-dependent
gauge parameter. 
}

\begin{document} 
\maketitle
\flushbottom

\section{Introduction}
\label{sec:intro}

%

%

Simultaneously with the modern diagrammatic approach to perturbative QED, 
in the early fifties Feynman developed a representation of the QED S-matrix in terms of 
first-quantised relativistic particle path integrals \cite{feynman:pr80,feynman:pr84}. 
For the simplest case, the one-loop effective action in scalar QED, this representation
can be written as

\bear
\Gamma_{\rm scal} [A] &=&
\int_0^{\infty}{dT\over T}\,{\rm e}^{-m^2T}
\int_P Dx 
\, {\rm e}^{-\int_0^T d\tau 
[ \fourth \dot x^2 + ie \dot x^{\mu}A_{\mu}(x) ]}\, .
\label{Gammascal}
\ear
Here $m$, $e$ and $T$ denote the mass, charge and proper-time of the loop scalar,
and $\int_{P}Dx $ the path integral over 
closed loops in (Euclidean) spacetime with periodicity $T$ in the proper-time
(the subscript `$P$' stands for ``periodic''). See appendix \ref{app-conv} for our 
conventions. 

Similarly, the tree-level scalar propagator in a background field is given by

\bear
D^{x'x}[A] &=&
\int_0^{\infty}
dT\,
\e^{-m^2T}
\int_{x(0)=x}^{x(T)=x'}
Dx\,
\e^{-\int_0^T d\tau\bigl[
\kinb
+ie\,\dot x\cdot A(x)
\bigr]}\, ,
\label{scalpropA}
\ear\no
where the propagation is from $x$ to $x'$. 
The external field in these formulas can be converted into photons by specialising it to
a sum of plane waves with definite momenta and polarisations, 

\bear
A^{\mu}(x) = \sum_{i=1}^N \varepsilon^{\mu}_i\,\e^{ik_i \cdot x}.
\label{Apw}
\ear
Each photon then gets effectively represented by a vertex operator (similar to those that appear in string perturbation theory)

\bear 
V_{\rm scal}[k,\varepsilon]=\int_0^Td\tau\, \veps\cdot\dot{x}(\tau)\,{\rm e}^{ik\cdot x (\tau)} \, ,
\label{vertop}
\ear
integrated along the scalar loop or line, with a coupling constant $(-ie)$ attached.
 Since in scalar QED any amplitude can be decomposed into scalar loops and/or lines adorned with any numbers
of external and internal photons\footnote{Here we disregard the quartic scalar vertex induced in scalar QED by the requirement of multiplicative renormalisability.}, 
starting from the formulas \eqref{Gammascal} and \eqref{scalpropA} one straightforwardly constructs
a path integral representation for the full scalar QED S-matrix \cite{feynman:pr80}.

To arrive at the analogous representation of the S-matrix in spinor QED, Feynman then simply adds on spin 
by the introduction of a ``spin factor'' ${\rm Spin}[x(\tau),A]$ in the path integral \cite{feynman:pr84}. For the closed loop case, this spin factor is

\bear
{\rm Spin}[x(\tau),A] &=& {\rm tr}_{\gamma} {\cal P}
\exp\biggl[{-i\frac{e}{4}[\gamma^{\mu},\gamma^{\nu}]
\int_0^Td\tau \, F_{\mu\nu}(x(\tau))}\biggr],
\label{defspinfactor}
\ear
where $F_{\mn}$ denotes the field strength tensor, ${\rm tr}_{\gamma}$ the Dirac trace, and ${\cal P}$ the path-ordering prescription.
Inserted into \eqref{Gammascal} it will (up to a global factor) 
convert the scalar loop effective action $\Gamma_{\rm scal} [A]$ into the spinor effective action $\Gamma_{\rm spin} [A]$: 

\bear
\Gamma_{\rm spin} [A] &=&
-\half \int_0^{\infty}{dT\over T}\,{\rm e}^{-m^2T}
\int_P D x(\tau)
\, {\rm Spin}[x(\tau),A]\,
\e^{-\int_0^T d\tau 
[ \fourth \dot x^2 + ie \dot x^{\mu}A_{\mu}(x) ]}
\, .
\label{Gammaspin}
\ear
This formalism, nowadays usually called the ``worldline formalism,'' was later on extended to other field theories 
(see \cite{41, UsRep} for a review and extensive bibliography). Nevertheless, it appears that for several decades it was considered mainly as of conceptual interest, rather than an alternative to the standard approach based on second quantisation and Feynman diagrams. In 1982 Affleck, Alvarez and Manton 
in a remarkable paper \cite{afalma} applied it to Schwinger pair creation in a constant field in scalar QED, even at the multiloop level, however, 
their ``worldline instanton'' formalism caught on only much later, after it was extended to spinor QED and non-constant fields in \cite{63,64}. 

This state of affairs changed only in the early nineties, when Strassler \cite{strassler1}, inspired by the seminal 
work of Bern and Kosower \cite{berkos} on the field theory limit of string amplitudes, developed an approach to the
calculation of such worldline path integrals that mimics string perturbation theory. The basic idea is quite simple, and was
germinally presented already in \cite{polyakovbook}: by suitable series expansions, the path integrals are reduced to Gaussian ones, 
and then evaluated by formal Gaussian integration as in a one-dimensional field theory, using appropriate ``worldline Green's functions.'' 

For example, in this formalism the calculation of the one-loop $N$-photon amplitude in scalar QED, starting from the
path integral representation \eqref{Gammascal}, proceeds as follows: after the above expansion of the interaction exponential, and
truncation to $N$th order, the amplitude is represented as

\bear
\Gamma_{\rm scal}(k_1,\varepsilon_1;\ldots ; k_N,\varepsilon_N) &=&
(-ie)^N 
\int_0^{\infty}
\frac{dT}{T}\,
\e^{-m^2T}
\int_P
Dx\,
\e^{-\int_0^T d\tau
\kinb}
\nonumber\\
&& \times
V_{\rm scal}[k_1,\varepsilon_1] V_{\rm scal}[k_2,\varepsilon_2]\cdots V_{\rm scal}[k_N,\varepsilon_N]\,.
 \nonumber\\
\label{DNpointscal}
\ear
The path integral is then split into an ordinary integral over the center-of-mass position
$x_0^{\mu}\equiv \frac{1}{T}\int_0^Td\tau \,x^{\mu}(\tau)$, 
and the path integral over the fluctuation
variable $q^{\mu}(\tau) \equiv x^{\mu}(\tau) - x_0^{\mu}$, subject to the nonlocal constraint

\bear
\int_0^T d\tau\,   q^{\mu} (\tau ) = 0 \, .
\label{nonlocal}
\ear
The integral over $x_0^{\mu}$ yields the global energy-momentum conservation factor 
${(2\pi)}^D\delta^D\bigl(\sum_{i=1}^N k_i\bigr)$. 
The path integral over $q^{\mu}(\tau)$ is already in Gaussian form, but to arrive at a closed-form evaluation
it is convenient, as in string theory, first to rewrite the photon vertex operator \eqref{vertop} in an exponential fashion as

\bear 
V_{\rm scal}^A[k,\varepsilon] =\int_0^Td\tau\, {\rm e}^{ik\cdot x(\tau)+\veps\cdot \dot{x}(\tau)}\Big\vert_{\veps}
=  {\rm e}^{ik\cdot x_0} \int_0^Td\tau\, {\rm e}^{ik\cdot q(\tau)+\veps\cdot \dot{q}(\tau)}\Big\vert_{\veps}\,,
\label{vertopexp}
\ear
where $\big\vert_{\veps}$ denotes the projection onto the terms linear in $\veps$. 
The path integration can then be computed by simply completing the square, leading to the following
master formula:

\begin{eqnarray}
\Gamma_{\rm scal}(k_1,\varepsilon_1;\ldots;k_N,\varepsilon_N)
&=&
{(-ie)}^N
{(2\pi )}^D\delta^D \big(\sum k_i\big)
{\dps\int_{0}^{\infty}}{dT\over T}
{(4\pi T)}^{-{D\over 2}}
\e^{-m^2T}
\prod_{i=1}^N \int_0^T 
d\tau_i
\nonumber\\
&&
\!\!\!\!\!\!\!\!\!\!\!\!\!\!
\times
\exp\biggl\lbrace\sum_{i,j=1}^N 
\Bigl\lbrack  \half G_{Bij} k_i\cdot k_j
-i\dot G_{Bij}\varepsilon_i\cdot k_j
+\half\ddot G_{Bij}\varepsilon_i\cdot\varepsilon_j
\Bigr\rbrack\biggr\rbrace
\Bigg\vert_{\veps_1\veps_2\cdots\veps_N}
\, .
\nonumber\\
\label{scalarqedmaster}
\end{eqnarray}
\no
Here we have introduced the Green function $G_B$, 

\bear
G_B(\tau,\tau') \equiv&
| \tau-\tau'| 
-{{(\tau-\tau')}^2\over T}\,, 
\label{defGB}
\ear
which (up to a constant that is irrelevant for our purposes here) is the Green's function for the second derivative operator adapted to the periodicity boundary
condition $q(T) = q(0)$ and the ``string-inspired''  constraint~\eqref{nonlocal},
and it is linked to the propagator of $q(\tau)$ by 

\bear
\langle q^\mu (\tau_{i}) q^{\nu}(\tau_{j}) \rangle = -G_{Bij}  \delta^\mn = - G_B(\tau_{i},\tau_{j}) \delta^\mn 
\label{qprop}
\ear
where we are abbreviating $G_B(\tau_i,\tau_j) \equiv G_{Bij}$ etc. The subscript `$B$' stands for ``bosonic'' (a ``fermonic'' Green
function $G_F$ will be introduced below).
Besides $G_B$ itself, also its first and second derivatives appear,

\bear
\dot G_B(\tau,\tau') &=& {\rm sign}(\tau-\tau') - 2 \frac{\tau-\tau'}{T} \, , \label{dotG}\\
\ddot G_B(\tau,\tau') &=& 2\delta(\tau-\tau') - \frac{2}{T} \, .\label{ddotG}
\ear
Here a `dot' always means a derivative with respect to the first variable.

The factor ${(4\pi T)}^{-{D\over 2}}$ comes from the free path integral:

\bear
{\dps\int} D q(\tau) \, \e^{-\int_0^T d\tau  \fourth \dot q^2} = {(4\pi T)}^{-{D\over 2}} \, .
\label{freepiq}
\ear
The notation $\big\vert_{\veps_1\veps_2\cdots \veps_N}$ means that the exponential should be expanded, and only the
terms linear in each of the polarisation vectors be kept. 

Although the master formula \eqref{scalarqedmaster}, as it stands, represents the off-shell one-loop $N$-photon amplitudes in scalar QED, 
it was originally derived by Bern and Kosower in the QCD context as a generating master expression from which to construct,
by purely algebraic means, parameter integral representations for the scalar, spinor and gluon loop contributions to the
on-shell $N$-gluon amplitudes \cite{berkos,berntasi,41}. 

In the scalar QED case, it is still straightforward to relate the parameter integrals resulting from the master formula to the ones obtained by 
a standard Feynman diagram calculation \cite{strassler1,berdun,41}. For any ordered sector of the $N$-fold proper-time integral 
$\int_0^Td\tau_1 \cdots \int_0^Td\tau_N$, 
the integrand can be identified with the Schwinger-parameter representation of the Feynman diagram with the corresponding ordering
of the photon legs, once the Schwinger parameters are identified with the differences of adjacent proper-time variables. The quartic
seagull vertex in this correspondence is presented by the delta function contained in $\ddot G_B$, equation \eqref{ddotG}.    
Despite this direct correspondence, the master formula is extremely useful for its compactness, and for combining into one integral
all the Feynman diagrams with different orderings of the $N$ photons. 
Although the latter property may not appear significant at the
one-loop level, when the $N$-photon amplitudes are used as building blocks for multiloop amplitudes it allows one to write down
highly nontrivial integral representations combining Feynman diagrams of different topologies, that would be hard to find using the standard 
formalism \cite{15,41}. 

Moreover, the representation of the integrand in terms of worldline Green's functions that are adapted to the periodic boundary conditions 
makes it possible to improve it by integration by parts (`IBP'), without generating boundary terms. An essential element of the original string-based approach by Bern and Kosower
cited above was the discovery that IBP could be used to eliminate all second derivatives $\ddot G_{Bij}$. In this way they obtained an integrand for the $N$ - gluon amplitude
where the prefactor of the exponential is written purely in terms of $\dot G_{Bij}$, and which offered the possibility, based on worldsheet supersymmetry,
to pass from the scalar to the spinor to gluon loop by applying simple 
pattern-matching rules to the integrand. Those involve the `$\tau$-cycles' 
$\dot G_{Bi_1i_2}\dot G_{Bi_2i_3}\cdots \dot G_{Bi_ni_1}$.

Later, Strassler \cite{strassler2} studied this IBP procedure in more detail for the case of the off-shell photon amplitudes, and found that it bears also an interesting relation to 
gauge invariance: a $\tau$ - cycle always appears multiplied by a corresponding `Lorentz-cycle', defined by 

\bear
Z_2(ij)&\equiv&
\half {\rm tr}(f_if_j) = \veps_i\cdot k_j\veps_j\cdot k_i - \veps_i\cdot\veps_jk_i\cdot k_j\,;
\nonumber\\
Z_n(i_1i_2\ldots i_n)&\equiv&
{\rm tr}
\Bigl(
\prod_{j=1}^n
f_{i_j}\Bigr) 
\quad (n\geq 3),
\nonumber\\
\label{defZn}
\ear\no
where $f_i$ is the field strength tensor associated to the $i$th photon/gluon,

\bear
f_i^{\mu\nu} \equiv k_i^{\mu}\veps_i^{\nu} - \veps_i^{\mu}k_i^{\nu}\, .
\label{deff}
\ear
Thus the integrand after the IBP can be written in terms of ``bosonic bi-cycles''

\bear
\dot G_B(i_1i_2\ldots i_n) \equiv \dot G_{Bi_1i_2}\dot G_{Bi_2i_3}\cdots \dot G_{Bi_ni_1}Z_n(i_1i_2\ldots i_n)\, ,
\label{defbosonicbicycle}
\ear
and certain left-overs called ``tails'' \cite{strassler2,26,41,91}. 

Generalising the master formula \eqref{scalarqedmaster} to the spinor QED case is a much less obvious task, and requires some
preliminary steps. For starters, we need to remove the path ordering implied in the definition of the Feynman spin factor ${\rm Spin}[x(\tau),A]$, equation \eqref{defspinfactor}.
This can be done using the following well-known identity, which represents the spin factor in terms of an auxiliary path integral over
Grassmann worldline fields $\psi^{\mu}(\tau)$:

\bear
{\rm tr}_{\gamma} {\cal P}
\exp\biggl[{-i\frac{e}{4}[\gamma^{\mu},\gamma^{\nu}]
\int_0^Td\tau \, F_{\mu\nu}(x(\tau))}\biggr]
=
{\dps\int_A} D\psi
\, \e^
{
-\int_0^Td\tau\,
(
\half\psi_{\mu}\dot\psi^{\mu}
-ie\psi^{\mu}F_{\mu\nu}\psi^{\nu}
)
}\,.
\nonumber\\
\label{idspinfactor}
\ear
Here the subscript `$A$' means anti-periodicity, $\psi^{\mu}(0) + \psi^{\mu}(T) =0$ which implements the Dirac trace. 
Apart from the removal of the path ordering, this replacement also leads to the appearance of a ``worldline supersymmetry''
between the $x$ and $\psi$ fields, 

\bear
\delta x^{\mu} &=& -2 \zeta\psi^{\mu}\,; \nonumber\\
\delta \psi^{\mu} &=& \zeta \dot x^{\mu}\,, \nonumber\\
\label{susy}
\ear
with a constant Grassmann parameter $\zeta$. Although this supersymmetry is broken by the boundary conditions,
its existence has far-reaching consequences in the worldline formalism \cite{strassler1,15,41}.  

After this replacement, one can proceed as in the scalar case, and find the following generalisation of \eqref{DNpointscal}:
\bear
\Gamma_{\rm spin}(k_1,\varepsilon_1;\ldots ; k_N,\varepsilon_N) &=&
-\half (-ie)^N 
\int_0^{\infty}
\frac{dT}{T}\,
\e^{-m^2T}
{\dps\int_P}
Dx\,
\e^{-\int_0^T d\tau
\kinb}
{\dps\int_A} D\psi
\, \e^
{-\int_0^Td\tau\, \half\psi\cdot\dot\psi}
\nonumber\\
&& \times
V_{\rm spin}[k_1,\varepsilon_1] V_{\rm spin}[k_2,\varepsilon_2]\cdots V_{\rm spin}[k_N,\varepsilon_N]
\, .
 \nonumber\\
\label{DNpointspin}
\ear
Here the photon vertex operator now takes the form

\bear
 V_{\rm spin}[k,\varepsilon] &\equiv &  \int_0^Td\tau\, \bigl[ \varepsilon\cdot \dot x(\tau)+2i \varepsilon\cdot \psi(\tau) k\cdot\psi(\tau)\bigr] \,\e^{ik\cdot x(\tau)}
 \nonumber\\
 &=& 
  \int_0^Td\tau\, \bigl[ \varepsilon\cdot \dot x(\tau) -
  i \psi(\tau)\cdot f\cdot\psi(\tau)\bigr] \,\e^{ik\cdot x(\tau)}
 \, . \nonumber\\
\label{defVspin}
\ear
Again the path integral \eqref{DNpointspin} is Gaussian, so that the only new information required for its
evaluation is the Green function for the Grassmann path integral. This one is simply $G_F(\tau,\tau') \equiv {\rm sign}(\tau-\tau')$,
and relates to the propagator of the $\psi$ field by

\bear
\langle \psi^{\mu}(\tau)\psi^{\nu}(\tau')\rangle = \half G_F(\tau,\tau') \delta^\mn\, .
\label{psiprop}
\ear
However, to arrive at a closed-form evaluation some rewriting is still necessary. 
This could be done in various ways, but we find it convenient to use the $N=1$ worldline superspace formalism \cite{strasslerthesis,41}:
we introduce a Grassmann super-partner $\theta$ for the proper-time $\tau$, and use it
to combine the worldline fields $x^{\mu}$ and $\psi^{\mu}$ into a superfield 

\begin{eqnarray}
X^{\mu}(\tau) &\equiv& x^{\mu}(\tau) + \sqrt 2\,\theta\psi^{\mu}(\tau)  \, .
\label{defX}
\ear
Introducing also $Q^{\mu} \equiv X^{\mu} - x^{\mu}_0$, and the super derivative 

\bear
D &\equiv & {\partial\over{\partial\theta}} - \theta{\partial\over{\partial\tau}} \label{defD}
\ear
we can then rewrite the kinetic term as

\bear
\int d\tau \left ( \kinq + \half \psi\cdot \dot\psi \right ) =  - \fourth \int d\tau \int d\theta\,  Q D^3 Q \, ,
\ear
(where $\int d\theta\theta = 1$), and the vertex operator \eqref{defVspin} in
a way analogous to the scalar case \eqref{vertop},

\bear
V_{\rm spin}[k,\varepsilon]  =  \int_0^Td\tau \int d\theta \,\varepsilon\cdot DQ \e^{ik\cdot X}
=
\e^{ik\cdot x_0} \int_0^Td\tau \int d\theta \, \e^{ik\cdot Q + \varepsilon\cdot DQ}\Big\vert_{\veps}
\, .
 \nonumber\\
\label{vertexsuper}
\ear
The double path integral in \eqref{DNpointspin} is then ready for a formal Gaussian integration, which leads to
the following master formula:

\begin{eqnarray}
\Gamma_{\rm spin}
(k_1,\varepsilon_1;\ldots;k_N,\varepsilon_N)
&=&
-2^{\frac{D}{2}-1}
{(-ie)}^N
{(2\pi )}^D\delta^D \big(\sum k_i\big)
{\dps\int_{0}^{\infty}}{dT\over T}
{(4\pi T)}^{-{D\over 2}}\e^{-m^2T}
\nonumber\\&&
\!\!\!\!\!\!\!\!\!\!\!\!\!\!\!\!\!\!\!\!
\!\!\!\!\!\!\!\!\!\!\!
\!\!\!\!\!\!\!\!\!\!\!\!\!\!\!\!\!\!\!\!
\!\!\!\!\!\!\!\!\!\!\!\!\!\!\!\!\!\!\!\!
\times
\prod_{i=1}^N \int_0^T 
d\tau_i
\int
d\theta_i
\exp\biggl\lbrace
\sum_{i,j=1}^N
\Big[
\half\widehat G_{ij} k_i\cdot k_j
+iD_i\widehat G_{ij}\varepsilon_i\cdot k_j
+\half D_iD_j\widehat G_{ij}\varepsilon_i\cdot\varepsilon_j\Big]
\biggr\rbrace
\Bigg\vert_{\varepsilon_1\ldots\varepsilon_N}\, .
\nonumber\\
\label{supermaster}
\end{eqnarray}
\no
Here we have introduced the super Green's function $\widehat G$, which combines $G_B$ and $G_F$:

\bear
\langle Q^{\mu}(\tau,\theta)Q^{\nu}(\tau',\theta')\rangle &=& - \widehat G(\tau,\theta;\tau',\theta')\delta^\mn, \nonumber\\
\widehat G(\tau,\theta;\tau',\theta') &\equiv& G_B(\tau,\tau') + \theta\theta' G_F(\tau,\tau')\, , \nonumber\\
\label{superpropagator}
\ear
which satisfies the Green equation in superspace $\frac{1}{2}D^{3}\widehat{G}(\tau, \theta; \tau^{\prime}, \theta^{\prime}) = \delta(\tau - \tau^{\prime})\delta(\theta - \theta^{\prime})$.

In the determination of the absolute sign of the amplitude, besides the $\theta_i$ and $d\theta_i$ also the $\veps_i$ have to be treated as Grassmann variables, 
and anticommuted into the standard ordering $\varepsilon_1\ldots\varepsilon_N$ at the end
(after the determination of the sign, the polarisation vectors turn into ordinary commuting quantities again, of course).
Our convention for the ordering of the $\theta$ integrals is $\int d\theta_1   \cdots  \int d\theta_N \theta_N \cdots \theta_1 =1$.
A factor of $2^{\frac{D}{2}}$ comes from the free $\psi$ path integral (which just counts the spin degrees of freedom). 
Here we assume that $D$ is even.

The parameter integrals resulting from the expansion of this master formula correspond to the Schwinger parameter integrals obtained by
Feynman diagrams in the same way as described above for scalar QED, however the comparison has to be done not with
the usual first-order Dirac formalism, but with the less familiar second-order formulation of spinor QED \cite{feygel,hostler,berdun,morgan, espin}. 
Its Feynman rules (see \cite{morgan}) are, up to global factors for statistics and degrees of freedom, the ones for scalar QED
with the addition of a third vertex due to the spin factor, involving $\sigma^{\mu\nu}\equiv {1\over 2}[\gamma^{\mu},\gamma^{\nu}]$.
We display them in Figure \ref{fig:secondorderrules}.

\begin{figure}[htbp]
\begin{center}
 \includegraphics[width=0.5\textwidth]{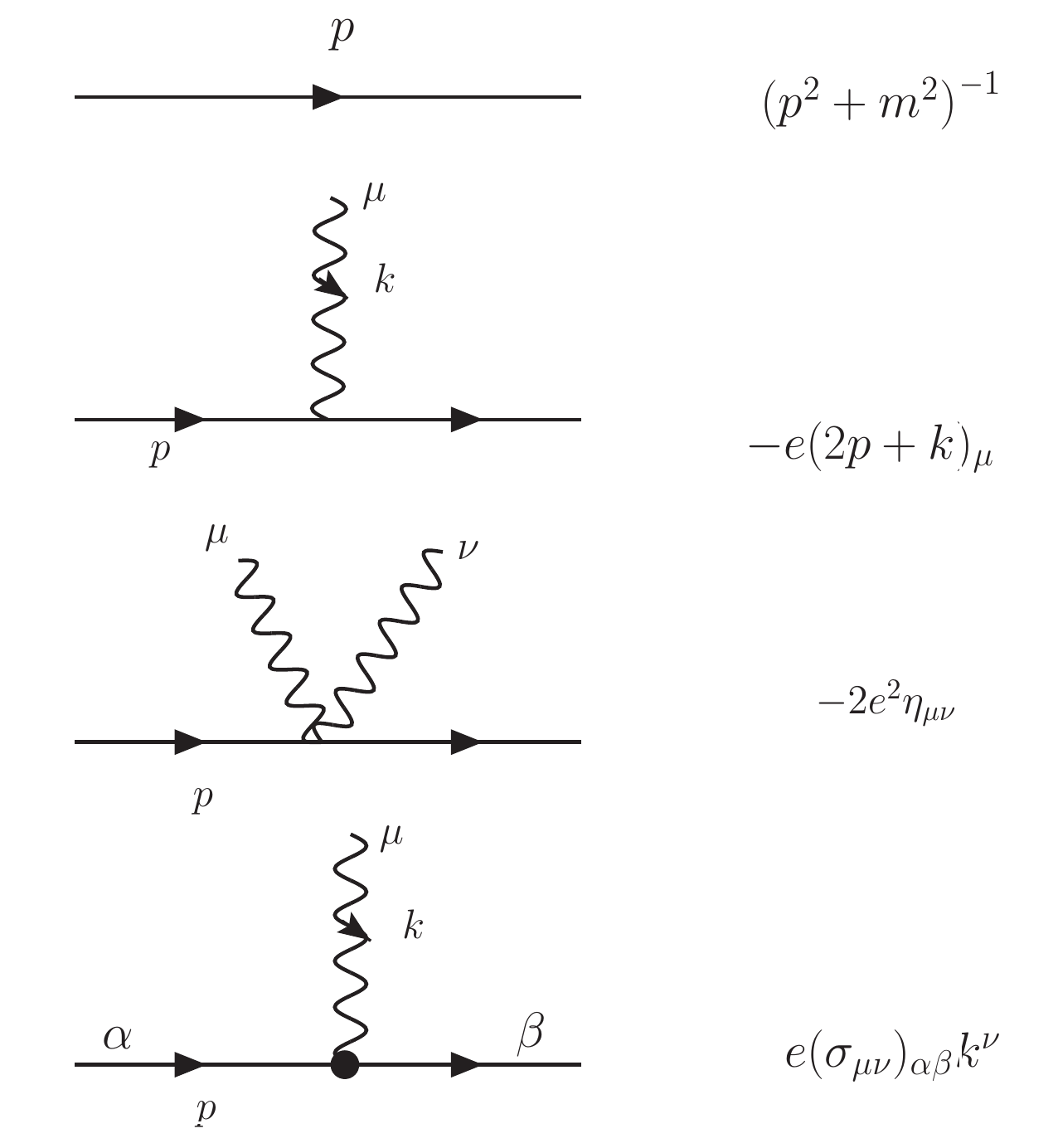}
\caption{Feynman rules for spinor QED in the second-order formalism.}
\label{fig:secondorderrules}
\end{center}
\end{figure}
Of course the final results for physical amplitudes coincide with those of the better known first order formalism.

During the last two decades, these ``string-inspired'' representations have already found
a considerable number of applications in QED, both for the calculation of photon amplitudes \cite{strassler1,15,51}
and the effective action itself \cite{5,cadhdu,gussho1,gussho2,95}. They have been generalised to include constant external fields \cite{shaisultanov,17,18,40,51,52}
as well as finite temperature \cite{mckreb:therm1,mckreb:therm2,shovkovy:therm,sato:therm,mckeon:therm,venwir}. Their non-Abelian generalisation was used in the first calculation of the
one-loop five-gluon amplitudes \cite{bediko5glu}, a calculation of the non-Abelian heat-kernel coefficients 
to fifth order \cite{25}, a calculation of the two-loop effective Lagrangian for a constant $SU(2)$ background field \cite{sascza},
and very recently for obtaining gauge-invariant decompositions of the off-shell three- and four-gluon amplitudes \cite{92,105,106,WI-qcd}. 
In the non-Abelian case, it may be helpful to generate the particle color factor, and take care of the path ordering, by adding suitable auxiliary fields,
in the same way as Grassmann variables take care of the spin factor and path ordering in \eqref{idspinfactor} -- see, for example,
\cite{Balachandran:1976ya,Barducci:1976xq,Bastianelli:2013pta,Ahmadiniaz:2015xoa}. Further applications to QCD-related topics can be found in references~\cite{Mueller:2017lzw, Mueller:2017arw, Mueller:2019qqj}. 

Extensions to curved space~\cite{Bastianelli:2002fv} and quantum gravity~\cite{Bastianelli:2013tsa, Bastianelli:2019xhi} have also been considered, addressing in particular induced effective actions and graviton self-energies 
\cite{Bastianelli:2002qw, Bastianelli:2005vk, Bastianelli:2005uy}, QED in curved spaces 
\cite{Hollowood:2007ku}, gravitational corrections to the Euler-Heisenberg Lagrangians
\cite{Bastianelli:2008cu, Davila:2009vt} and related amplitudes \cite{Bastianelli:2012bz}, and
studies of one-loop photon-graviton conversion in strong magnetic fields \cite{Bastianelli:2004zp,Bastianelli:2007jv}.
The case of higher spin fields has also been approached using worldlines \cite{Bastianelli:2007pv, Bastianelli:2008nm,Corradini:2010ia,
Bastianelli:2012bn}, as has quantum field theory on non-commutative spaces \cite{Bonezzi:2012vr, Kiem:2001dm, NCU1, NCUN} and spaces with boundary~\cite{Bastianelli:2006hq, Bastianelli:2008vh, ConfinedScalar}.

However, with a few exceptions as in \cite{mckeon:ap224,mckreb:gamma,mckshe,karkto} and 
\cite{Casalbuoni:1974pj,Dai:2006vj,Holzler:2007xt, Dai:2008bh,Bonezzi:2018box,ahmadiniaz2020compton},
these applications have been restricted to processes involving only closed scalar or spinor loops,
not open lines. For the scalar QED case, Daikouji et al. \cite{dashsu} have obtained the following master formula, analogous to \eqref{scalarqedmaster}, for the scalar propagator
dressed with $N$ photons:

\bear 
D^{p'p}(k_1,\varepsilon_1;\cdots; k_N,\varepsilon_N)&=&(-ie)^N(2\pi)^D\delta^D\Big(p+p'+\sum_{i=1}^Nk_i\Big)\int_0^\infty dT\,{\rm e}^{-m^2T}\nonumber\\
&&\hspace{-1.5cm}\times \prod_{i=1}^N\int_0^Td\tau_i\,
{\rm e}^{-Tb^2+\sum_{i,j=1}^N[\Delta_{ij}k_i\cdot k_j-2i\ddel_{ij}\varepsilon_i\cdot k_j-\ddeld_{ij}\varepsilon_i\cdot \varepsilon_j]}\Big\vert_{\varepsilon_1\varepsilon_2\cdots\varepsilon_N}\, .\nonumber\\
\label{linemaster} 
\ear
Here we have introduced the vector

\bear
b \equiv p'+\frac{1}{T}\sum_{i=1}^N(k_i\tau_i-i\varepsilon_i),
\label{defb}
\ear
and a different worldline Green's function $\Delta(\tau,\tau')$ has been used for the $q$ propagator:

\bear
\langle q^\mu (\tau) q^{\nu}(\tau') \rangle &=& - 2 \Delta(\tau,\tau') \delta^\mn ,\nonumber\\
\Delta(\tau,\tau')&=&\frac{\vert\tau-\tau'\vert}{2}-\frac{\tau+\tau'}{2}+\frac{\tau\tau'}{T} 
\,. \nonumber\\
\label{defDelta} 
\ear 
Instead of the string-inspired boundary conditions \eqref{nonlocal}, this Green's function is adapted to Dirichlet boundary conditions, 

\bear
q^{\mu}(0) = q^{\mu}(T) = 0 \, .
\label{DBC}
\ear
These boundary conditions break the translation invariance in proper-time,
so that one now has to distinguish between derivatives with respect to the first and the second argument.
A convenient notation is \cite{basvan-book} to use left and right dots to indicate derivatives
with respect to the first and the second argument, respectively:

\bear 
\ddel(\tau,\tau')&=&\frac{\tau'}{T}+\frac{1}{2}{\rm
sign}(\tau-\tau')-\frac{1}{2}\,,\nonumber\\
\deld(\tau,\tau')&=&\frac{\tau}{T}-\frac{1}{2}{\rm
sign}(\tau-\tau')-\frac{1}{2}\,,\nonumber\\
\ddeld(\tau,\tau')&=&\frac{1}{T}-\delta(\tau-\tau')\,.
\nonumber\\
\label{derDelta}
\ear
We will also need the coincidence limits

\bear
\Delta(\tau,\tau)&=&\frac{\tau^2}{T}-\tau \, ;\non
\ddel(\tau,\tau)&=&\deld(\tau,\tau)=\frac{\tau}{T}-\half \, .\non
\ear 
Note that, apart from the different boundary conditions, the Green's functions $\Delta $ and $G_B$ differ also by a conventional factor of two
in their normalisation. Finally, since $\Delta$ is somewhat less convenient than $G_B$, 
it is sometimes useful to observe that the two are related by

\bear
2\Delta(\tau,\tau') = G_B(\tau,\tau') - G_B(\tau,0) - G_B(0,\tau') \, .
\label{relDeltaG}
\ear
The master formula \eqref{linemaster} represents the {\sl un-truncated} dressed propagator, that is the
sum of diagrams given in Fig. \ref{fig-multiphoton}, where the final scalar propagators at each end are included.
This technical point will play an important role in the following. The momenta $p, p', k_1,\ldots, k_N$ are all ingoing. 

\begin{figure}[h]
  \centering
   \includegraphics[width=0.8\textwidth]{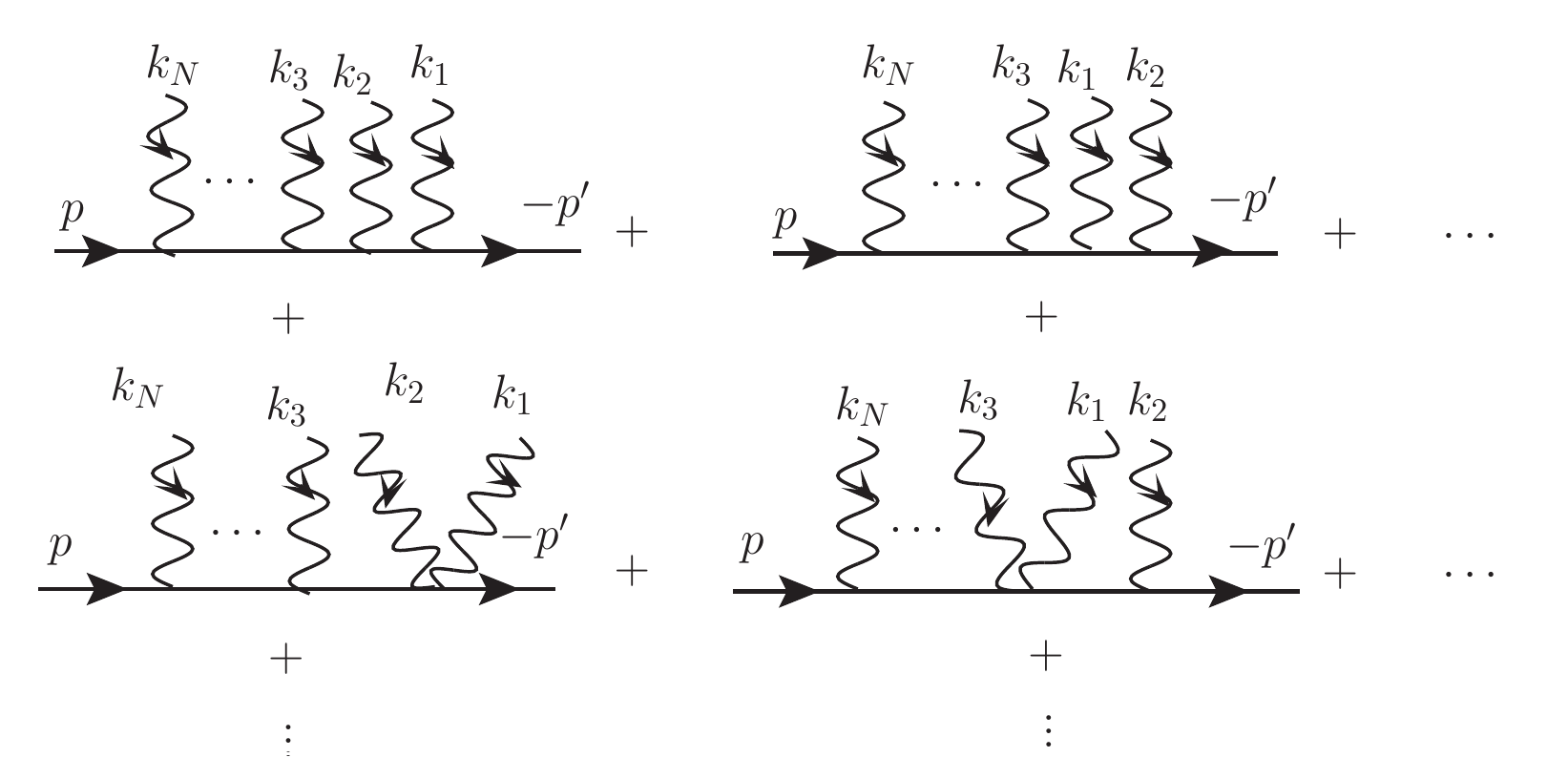}
\caption{Multi-photon Compton scattering diagram in scalar QED (we do not distinguish the propagator of the matter field between scalar or spinor QED, choosing to indicate both with a solid line). The seagull vertices in the second row are once again produced by the $\delta$-function in the second derivative of the open-line Green function $\Delta$.} 
\label{fig-multiphoton}
\end{figure}

In \cite{dashsu} it was obtained by a comparison with the Schwinger-parameter representation of the corresponding Feynman diagrams. The same formula 
has recently been rederived from the path integral representation \eqref{scalpropA} in \cite{102}. 

The worldline formalism has also been applied to the fermion line case \cite{mckreb:gamma,karkto}, but a Bern-Kosower type
master formula for the dressed propagator has not been derived so far. The purpose of the present paper is to 
solve this long-standing problem, obtain such a formula, and to demonstrate its usefulness as an alternative to 
the standard Feynman diagram formalism. 
We will start from the well-known second-order representation of the $x$-space Dirac propagator $S^{x'x}[A]$ in a Maxwell background, 
 
\bear
S^{x'x}[A] &=&
\bigl[m + i\s{D}'\bigr]
K^{x'x}[A]\, ,
\label{StoK}
\ear
where $\s D = \gamma^{\mu}D_{\mu}$, $D_{\mu} = \partial_{\mu} + ieA_{\mu} $ and\footnote{See appendix A for our conventions.}

\bear
K^{x'x}[A] &\equiv &
\Big\langle x'\Big|\Bigl[m^2- D_{\mu}D^{\mu} +{i\over 2}\, e \gamma^{\mu}\gamma^{\nu} F_{\mu\nu}\Bigr]^{-1} 
\Big| x \Big \rangle
\, .
\label{defK}
\ear
For this ``kernel'' function, we will then derive the following path integral representation:

\bear
K^{x'x}[A] &=&
\int_0^{\infty}
dT\,
\e^{-m^2T}
\e^{-\fourth \frac{(x-x')^2}{T}}
\int_{q(0)=0}^{q(T)=0}
Dq\,
\e^{-\int_0^T d\tau\bigl(
\kinq
+ie\,\dot q\cdot A 
+ie \frac{x'-x}{T}\cdot A
\bigr)}
\nonumber\\
&& \times\,  2^{-\frac{D}{2}}
{\rm symb}^{-1}
\int_{\psi(0)+\psi(T)=0
\hspace{-30pt}}D\psi
\, \e^
{-\int_0^Td\tau\,
\bigl[\half\psi_{\mu}\dot\psi^{\mu}-ieF_{\mu\nu}(\psi+\eta)^{\mu}(\psi+\eta)^{\nu}\bigl]
}\, .
\label{Kfin}
\ear\no
Here $\eta^{\mu}$ is an external Grassmann Lorentz vector, and the ``symbol map,'' {\it symb}, converts products of $\eta$'s 
into fully antisymmetrised products of Dirac matrices; we will discuss the details in section \ref{fermionpi} below.

Following this we perform the usual projection onto an $N$-photon background, and use the path integral representation \eqref{Kfin}
to derive master formulas for the $N$ - photon kernel $K$ both in configuration and in momentum space. 
Those master formulas, given later in \eqref{Ndressedmasterx} and \eqref{Ndressedmasterp}, are the central results of the paper. 

Returning from the kernel $K$ to the propagator itself, we will then also Fourier transform our starting identity \eqref{StoK} to momentum space. Projected on the $N$-photon sector, it turns into

\bear
S_N^{p'p} [\varepsilon_1,k_1;\ldots;\varepsilon_N,k_N]  &=& ({\s p'}+m) K_N^{p'p}[\varepsilon_1,k_1;\ldots;\varepsilon_N,k_N] 
\nonumber\\ &&
- e\sum_{i=1}^N{\s \varepsilon_i} K_{(N-1)}^{p'+k_i,p}[\varepsilon_1,k_1;\ldots;\hat\varepsilon_i,\hat k_i; \ldots\varepsilon_N,k_N] \, .
\nonumber\\
\label{SN}
\ear
Here in the second term the `hat' on $\varepsilon_i$ and $k_i$ means omission. 
We will work out this formula for $N=0,1,2$ to see how the equivalence to the standard formalism comes about in detail. 
As expected, the
scalar QED calculations are close to the standard field theory ones, while in the fermion
case nontrivial rearrangements have to be done to match the textbook Feynman diagram
calculations.


The organisation of this paper, the first part in a series of three, is as follows: as a warm-up, in section \ref{scalarqed} we shortly retrace the derivation of the scalar open line master formula
\eqref{scalarqedmasteropen} from the path integral representation \eqref{scalpropA}, following \cite{102}. 
In section \ref{fermionpi} we derive the worldline path integral representation \eqref{Kfin} of the kernel $K$ starting from field theory. 
Sections \ref{Kx} and \ref{Kp} contain the derivations of the 
configuration and momentum space master formulas, respectively.     
In section \ref{spinorbit} we provide a different derivation of the $N$-photon kernel that keeps track of the orbital and spin contributions to the basic interaction between the electron and the photon in the
underlying second-order formalism. 
We then move on from the $N$-photon kernel to the fully dressed electron propagator in section \ref{S}. 
We work out the cases $N=0,1,2$ and study the equivalence to the standard formalism, still off-shell, which 
happens in a quite non-obvious way. 
As a state-of-the-art application,
in section \ref{self energy} 
we recalculate the one-loop fermion self energy (in arbitrary dimension and covariant gauge). 
Section \ref{conc} offers our summary, and an outlook on future applications and generalisations.

There are four appendices: appendix \ref{app-conv} lists our conventions. 
Appendix \ref{app-intrinsic}
offers an alternative, more ``principled,''  derivation of the 
worldline path-integral representation of the electron propagator that,
contrary to the one given in the main text, minimises the use of field theory concepts. 
Appendix \ref{app-symb} is devoted to the representation of the Feynman spin factor in terms of a Grassmann path
integral.
Finally, in \ref{app-hypergeo} we prove a hypergeometric identity that we use in section \ref{self energy} to simplify
our result for the fermion self energy.  

The forthcoming second part of this series will be devoted to the use of the 
master formulas derived here for on-shell calculations such as cross-sections, the third part to 
the inclusion of a constant electromagnetic background field. 

\section{The dressed propagator in scalar QED}
\label{scalarqed}
In this section, we wish to derive the master formula, eventually given in \eqref{scalarqedmasteropen}, for the dressed propagator, and discuss some of its properties. 

\subsection{Derivation of the scalar master formula}
\label{derivescalarmaster}

Starting from \eqref{scalpropA}, and proceeding as in the closed-loop case, we get a representation of this propagator 
in terms of the photon vertex operator \eqref{vertop} analogous to \eqref{DNpointscal}:

\bear 
D_N^{x'x}(k_1,\varepsilon_1;\cdots; k_N,\varepsilon_N)&=&
(-ie)^N 
\int_0^{\infty}
dT
\e^{-m^2T}
\int_{x(0)=x}^{x(T)=x'}
Dx\,
\e^{-\int_0^T d\tau
\kinb}
\nonumber\\
&& \times
V_{\rm scal}[k_1,\varepsilon_1] V_{\rm scal}[k_2,\varepsilon_2]\cdots V_{\rm scal}[k_N,\varepsilon_N]
\, .
 \nonumber\\
 \label{linevertop} 
\ear
Shifting the path integration variable as

\bear
x(\tau) = x_0(\tau) + q(\tau)\,,
\label{shift}
\ear
where  $x_0$ is the straight-line trajectory

\bear
x_0(\tau) = x + (x'-x) \frac{\tau}{T} \,,
\label{defx0}
\ear
reduces the boundary conditions to Dirichlet boundary conditions,

 \bear
 q(0) = q(T) = 0 \, .
 \label{dirichlet} 
 \ear
Rewriting the photon  vertex operator as in \eqref{vertopexp}, \eqref{linevertop} becomes 

\bear 
D_N^{x'x}(k_1,\varepsilon_1;\cdots; k_N,\varepsilon_N)&=&
(-ie)^N\intT\, {\rm e}^{-\frac{1}{4T}(x-x')^2}\int_{q(0)=0}^{q(T)=0}
Dq\,\eTq\nonumber\\
&\times&
\prod_{i=1}^N  \int_0^T d\t i\,
{\rm e}^{\sum_{i=1}^N\big(\veps_i\cdot \frac{(x'-x)}{T}+\veps_i\cdot\dot{q}(\t i)+ik_i\cdot (x'-x)\frac{\t i}{T} +ik_i\cdot x+ik_i\cdot q(\t i)\big)}
\multilin
\, .
 \nonumber\\
 \label{linevertoprewrite} 
\ear
The path integral can now be performed by formal Gaussian integration using the worldline Green function $\Delta(\tau,\tau')$, leading to

\bear
D_N^{x'x}(k_1,\varepsilon_1;\cdots; k_N,\varepsilon_N)
&=&(-ie)^N\intT\, {\rm e}^{-\frac{1}{4T}(x-x')^2}\big(4\pi
T\big)^{-\frac{D}{2}}\nonumber\\ && \hspace{-3.5cm}
\times
\prod_{i=1}^N\int_0^Td\tau_i\,
{\rm e}^{\sum_{i=1}^N\big(\veps_i\cdot \frac{(x'-x)}{T}+ik_i\cdot (x'-x)\frac{\t i}{T}+ik_i\cdot x\big)}\, {\rm e}^{\sum_{i,j=1}^N\big[\Delta_{ij}\kk
ij-2i\ddel_{ij}\veps_i\cdot k_j-\ddeld_{ij}\epseps
ij\big]}
\multilin
\,.\nonumber\\
\label{bk-like-x}
 \ear
 (the free path integral normalisation \eqref{freepiq} holds for Dirichlet boundary conditions as well). 
 Finally, we also Fourier transform the scalar legs of the master
formula in equation ~(\ref{bk-like-x}) to momentum space:
\bear
D_N^{p'p}(k_1,\varepsilon_1;\cdots; k_N,\varepsilon_N)
&\equiv&
\int d^Dx\int d^Dx'\, {\rm e}^{ip\cdot x+ip'\cdot x'}\,
D^{x'x}(k_1,\varepsilon_1;\cdots; k_N,\varepsilon_N)\,.
\label{fourier}
\ear
After  a change of variables from $x,x'$ to $x_\pm$, defined by

\bear
x_+ &=& \frac{1}{2}(x+x')\,, \nonumber\\
x_- &=& x'-x\,, \nonumber\\
\label{cov}
\ear
the integral over $x_+$ produces the delta-function for the total
conservation of  energy-momentum:

 \bear
\hspace{-1.5em} D_N^{p'p}(k_1,\varepsilon_1;\cdots; k_N,\varepsilon_N)
&=&(-ie)^N(2\pi)^D\delta^D\Big(p+p'+\sum_{i=1}^N k_i\Big)
\int_0^\infty dT\,{\rm e}^{-m^2T}(4\pi T)^{-\frac{D}{2}}
\int d^Dx_- \,{\rm e}^{-\frac{1}{4T}x_-^2}
\non &&
\hspace{-1.5em}\hspace{-2cm} \times 
\prod_{i=1}^N\int_0^Td\tau_i\,
{\rm e}^{ix_-\cdot
\big(p'+ \frac{1}{T}\sum_{i=1}^N (k_i\t i -i \varepsilon_i ) \big)}\,
 {\rm e}^{\sum_{i,j=1}^N\big[\Delta_{ij}\kk
ij-2i\ddel_{ij}\varepsilon_i\cdot k_j-\ddeld_{ij}\epseps
ij\big]}
\multilin
\,.\non
\label{e1} \ear
Performing also the $x_-$ integral, one arrives at the momentum space master formula given in the introduction, \eqref{linemaster}. 

Let us also introduce some more notation here.
The result of expanding out the exponential factor in (\ref{linemaster}) will be named
$(-i)^N \bar P_N\,\e^{(\cdot)}$, namely

\bear 
(-i)^N \bar P_N\,\e^{(\cdot)}=
{\rm e}^{-Tb^2+\sum_{i,j=1}^N\big[\Delta_{ij}k_i\cdot k_j-2i\ddel_{ij}\varepsilon_i\cdot k_j-\ddeld_{ij}\varepsilon_i\cdot \varepsilon_j\big]}\Big\vert_{\varepsilon_1\varepsilon_2\cdots\varepsilon_N}\, ,\nonumber\\
\ear
where now 

\bear 
\e^{(\cdot)}\equiv {\rm e}^{-Tb_0^2+\sum_{i,j=1}^N \Delta_{ij}k_i\cdot k_j}\,,
\label{expred}
\ear
and

\bear
b_0 \equiv b\vert_{\veps_1= \cdots = \veps_N =0} = p'+\frac{1}{T}\sum_{i=1}^N k_i\tau_i
\label{defb0}
\ear
(the `bar' on $P_N$ is to distinguish it from the corresponding quantity for the closed loop \cite{41}).

\subsection{Off-shell IBP}
\label{ibpoffshell}

One of the advantages of this master formula's encoding of the usual Feynman-parameter integrals
in terms of the worldline Green function $\Delta$ is that IBP can be used to remove the second derivative $\ddeld$, and thus the
seagull vertex. This homogenises the integrand and leads to the automatic appearance of field strength tensors, which again
can be arranged into bi-cycles. Those now take the form (compare \eqref{defbosonicbicycle})

\bear
\ddel(i_1i_2\ldots i_n) \equiv \ddel_{i_1i_2}\ddel_{i_2i_3}\cdots \ddel_{i_ni_1}Z_n(i_1i_2\ldots i_n)  \quad (n\geq 2)\, .
\label{defbosonicbicyclemod}
\ear
The same IBP algorithm as in the closed-loop case \cite{26,41} can be used, and non-vanishing boundary terms are still not generated
due to $\Delta\targ$ obeying Dirichlet boundary conditions. The resulting integrand will be called $(-i)^N \bar Q_N\, \e^{(\cdot)}$.  
$\bar Q_N$ in general will contain both $\deld\targ$ and $\ddel\targ$, but we will standardise it using the identity $\deld \targ = \ddel (\tau',\tau)$
to trade the former for the latter throughout. 

We will call this IBP algorithm ``Off-shell IBP'' because it is designed not to generate any boundary terms.
Below we will define an alternative IBP procedure that seems preferable in the on-shell case. 

Let us work out the integrand for $N=1,2$. For $N=1$, the expansion of the exponential factor in (\ref{linemaster}) yields

\bear
{\bar P_1} &=& 2\ddel_{11}\veps_1\cdot k_1 - 2\veps_1\cdot b_0 \, . \nonumber\\
\label{barP1}
\ear
Here there are no second derivatives yet, so $\bar Q_1=\bar P_1$. 

\noindent
For $N=2$ we find

\bear
{\bar P}_2 &=& 4  (\ddel_{12}\veps_1\cdot k_2 + \ddel_{11}\veps_1\cdot k_1-\veps_1\cdot b_0) 
(\ddel_{21}\veps_2\cdot k_1 + \ddel_{22}\veps_2\cdot k_2 - \veps_2\cdot b_0) \nonumber\\
&& - \bigl(\frac{2}{T}-2 \ddeld_{12}\bigr)\veps_1\cdot\veps_2 \,.
\label{barP2}
\ear
Here the last term asks for IBP, which transforms it as

\bear
2 \ddeld_{12}\veps_1\cdot\veps_2  \longrightarrow &-& \Bigl( 4 \ddel_{12} \ddel_{21}k_1\cdot k_2 
+2 \ddel_{11} \ddel_{21}k_1^2
+2 \ddel_{12} \ddel_{22}k_2^2 
-2 \ddel_{12}k_2\cdot b_0 \nonumber \\
&&- 2\ddel_{21}k_1\cdot b_0\Bigr)
\veps_1\cdot\veps_2\, , \nonumber\\
\ear 
where we have used the identity $\partder{}{\tau_i}\Delta_{ii} = 2\ddel_{ii}$.
Sorting according to cycles, ${\bar Q}_2$ can be written as

\bear
{\bar Q}_2 &=& 4\ddel(12) 
 +4  \ddel_{12}\veps_1\cdot k_2
 (\ddel(2) - \veps_2\cdot b_0)
 +4  (\ddel(1)-\veps_1\cdot b_0) 
\ddel_{21}\veps_2\cdot k_1 
\nonumber\\ &&
 +4  (\ddel(1)-\veps_1\cdot b_0) 
(\ddel(2) - \veps_2\cdot b_0)
\nonumber\\ &&
- \Bigl(\frac{2}{T}
+2 \ddel_{11} \ddel_{21}k_1^2
+2 \ddel_{12} \ddel_{22}k_2^2 
-2 \ddel_{12}k_2\cdot b_0 - 2\ddel_{21}k_1\cdot b_0
\Bigr)\veps_1\cdot\veps_2 \, .
\nonumber\\
\ear
where we used the following compact notation
$$\ddel(i)\equiv \ddel_{ii}\epsk ii.$$

\subsection{The QED Ward identity}

The amplitude $D_N^{p'p}(k_1,\varepsilon_1;\cdots; k_N,\varepsilon_N)$
should fulfill the QED Ward identity, i.e. replacing any 
\begin{equation}
\varepsilon_i \to k_i
\label{ward}
\end{equation}
should give something that does not contribute on-shell. This property is not obvious
from the master formulas \eqref{e1} or \eqref{linemaster}, but is easily seen in the
path-integral representation \eqref{linevertop}. The replacement \eqref{ward} turns the
vertex operator $V_{\rm scal}[k_i,\varepsilon_i]$ into 

\bear
V_{\rm scal}[k_i,k_i] = \frac{1}{i} \int_0^Td\tau_i \frac{d}{d\tau_i} \e^{ik_i\cdot x(\tau)}
= -i (\e^{ik_i\cdot x(T)} -  \e^{ik_i\cdot x(0)})
=  -i (\e^{ik_i\cdot x'} -  \e^{ik_i\cdot x})\,.
\nonumber\\
\ear
Under the Fourier transformation \eqref{fourier}, the first term in brackets will
change the denominator of the right-most scalar propagator (in the conventions
of fig. \ref{fig-multiphoton}) from $p'^2+m^2$ to $(p'+k_i)^2+m^2$, while
the second term changes the denominator of the left-most scalar propagator from
$p^2+m^2$ to $(p+k_i)^2+m^2$. Thus neither term conserves the double pole 
that by the LSZ theorem is necessary for contributing to the on-shell matrix element.

We can take advantage of the Ward identity to achieve manifest gauge invariance at the
integrand level. Namely, let us choose for each $k_i$ a ``reference vector'' $r_i$ such that
$k_i\cdot r_i \ne 0$, and define the modified vertex operator

\bear
V_{\rm scal}[k,\veps,r] &\equiv& V_{\rm scal}[k,\veps] 
+ i  \frac{\veps\cdot r}{k \cdot r} \int_0^Td\tau \frac{d}{d\tau} \e^{ik\cdot x(\tau)}\non
&=&
\int_0^Td\tau \frac{r\cdot f\cdot {\dot x}}{r\cdot k} \e^{ik\cdot x(\tau)}
= \int_0^Td\tau \e^{ik\cdot x(\tau)+\frac{r\cdot f\cdot {\dot x}}{r\cdot k}}\Big\vert_{f} \,.
\label{defVertopmod}
\ear
Plugging this back into (\ref{linevertop}) and Fourier transforming to momentum space the on-shell version of the master formula for the dressed propagator in terms of the field strength tensors is given by  
 \bear
\hspace{-4em} &&\mathcal{M}^{p'p}(k_1,f_1;\cdots k_N,f_N)\equiv (p'^2+m^2)D_N^{p'p}(k_1,f_1;\cdots; k_N,f_N)(p^2+m^2)\non
\hspace{-4em}&&=(-ie)^N(2\pi)^D\delta^D\Big(p+p'+\sum_{i=1}^N k_i\Big)
\int_0^\infty dT\,{\rm e}^{-m^2T}(4\pi T)^{-\frac{D}{2}}
\int d^Dx_- \,{\rm e}^{-\frac{1}{4T}x_-^2}
\non 
\hspace{-6em} && \hspace{-1em}
\times 
\prod_{i=1}^N\int_0^Td\tau_i\,
{\rm e}^{ix_-\cdot
\big(p'+ \frac{1}{T}\sum_{i=1}^N (k_i\t i +i\frac{f_i\cdot r_i}{r_i\cdot k_i}  ) \big)}\,
 {\rm e}^{\sum_{i,j=1}^N\big[\Delta_{ij}\kk
ij-2i\ddel_{ij}\frac{r_i\cdot f_i\cdot k_j}{r_i\cdot k_i}-\ddeld_{ij}\frac{r_i\cdot f_i\cdot f_j\cdot r_j}{r_i\cdot k_i\,r_j\cdot k_j}\big]}\Big\vert_{f_1f_2\cdots f_N}
\,,\non
\label{mas-fs} \ear
which will be discussed more in the forthcoming paper. 
Retracing the derivation of the master formula \eqref{scalarqedmaster} with this modified vertex operator, we arrive at the ``covariantised Bern-Kosower master formula'' 
(henceforth we usually omit the
global energy-momentum conservation factor)

\begin{eqnarray}
\Gamma_{\rm scal}[k_1,\varepsilon_1;\ldots;k_N,\varepsilon_N]
&=&
{(-ie)}^N
{\dps\int_{0}^{\infty}}{dT\over T}
{(4\pi T)}^{-{D\over 2}}
\e^{-m^2T}
\prod_{i=1}^N \int_0^T 
d\tau_i
\nonumber\\
&&
\hspace{-90pt}
\times
\exp\biggl\lbrace\sum_{i,j=1}^N 
\Bigl\lbrack  \half G_{Bij} k_i\cdot k_j
-i\dot G_{Bij}{r_i\cdot f_i\cdot k_j\over r_i\cdot k_i}
-\half  \ddot G_{Bij} 
{r_i\cdot f_i\cdot f_j\cdot r_j\over r_i\cdot k_i \, r_j\cdot k_j}
\Bigr\rbrack\biggr\rbrace
\Bigg\vert_{f_1f_2\ldots f_N}\, .
\nonumber\\
\label{scalarqedmastercov}
\end{eqnarray}
\no
In \cite{91} this version of the master formula was obtained by IBP at the parameter integral level and
called the ``R-representation.''
Note that it reduces to the original master formula (\ref{scalarqedmaster}) if $r_i\cdot\veps_i=0$ for all $i$.

\subsection{Alternative forms of the master formula} \label{altmf}

Finally, let us also give two alternative forms of the master formula \eqref{linemaster}. Writing the worldline Green function explicitly, and
taking advantage of some cancellations in the exponent, one can rewrite it as 

\bear
&&D_N^{p'p}(k_1,\veps_1;\cdots;k_N,\veps_N) = (-ie)^N
\int_0^\infty dT\, {\rm e}^{-T(m^2+p'^2)}\nonumber\\
&\times&\int_0^T\prod_{i=1}^{N} d\tau_i\, {\rm e}^{\sum_{i=1}^N(-2k_i\cdot p'\tau_i+2i\veps_i\cdot p')+\sum_{i,j=1}^N\big[\bigl(\frac{\vert \tau_i-\tau_j\vert}{2}-\frac{\tau_i+\tau_j}{2}\bigr)k_i\cdot k_j-i({\rm sign}(\tau_i-\tau_j)-1)\veps_i\cdot k_j+\delta(\tau_i
-\tau_j)\veps_i\cdot\veps_j\big]}
\Big\vert_{\veps_1\veps_2\cdots \veps_N}\,. \nonumber\\
\label{scalarqedmasteropen}
\ear
And this can be written even more compactly at the expense of introducing some more notation.
Namely, defining 

\bear
k_0 &\equiv& p'\, , \quad
k_{N+1} \equiv p\,,\quad
\tau_0 \equiv  T \, ,\quad
 \tau_{N+1}  \equiv  0 \, ,\quad
 \varepsilon_0 \equiv  0 \, , \quad
 \varepsilon_{N+1} \equiv 0 \, , \nonumber\\
 \label{defKextension} 
\ear 
we can, using energy-momentum conservation in the exponent, arrive at the following form:
\bear 
D_N^{p'p}(k_1,\veps_1;\cdots;k_N,\veps_N) &=& (-ie)^N
\int_0^\infty dT\, {\rm e}^{-m^2T}
\nonumber\\ &\times&
\int_0^T\prod_{i=1}^{N} d\t i\,
{\rm e}^{\sum_{i,j=0}^{N+1}\big[\frac{1}{2}\vert \tau_i-\tau_j\vert
k_i\cdot k_j -i {\rm sign}(\tau_i-\tau_j)\varepsilon_i\cdot
k_j+\delta (\tau_i-\tau_j)\epseps ij \big]}
\multilin
\, .
\nonumber\\ \label{master-dss} \ear
It is this form of the momentum space master formula that was previously 
obtained by Daikouji {\em et al.}~\cite{dashsu} by a direct
comparison with the corresponding Feynman-Schwinger parameter
integrals and later in \cite{102} using the worldline formalism. It turns out that the above formula has the advantage of leading to manifest
worldline Poincar\`e invariance on the mass-shell of the scalar particle \cite{Ahmadiniaz:2015xoa}.  
This  provides a worldline analogue of the well-known fact that in string theory the worldsheet theory
becomes conformally invariant only if all vertex operator insertions are on-shell.

\section{Path integral representation of the electron propagator in an Abelian background field}
\label{fermionpi}

Contrary to the scalar case, there are various routes to obtain a worldline path integral representation of the fermion propagator in a Maxwell
background. In this section, we will present a field-theory based construction that essentially follows \cite{fragit}, delegating some
technical details to appendix \ref{app-symb}. The same representation is rederived in appendix \ref{app-intrinsic} from an intrinsic worldline point of view.

The most specific feature of the method presented here, is the use of ``Weyl symbols'', defined in \eqref{defsymb} below, to represent fermionic operators \cite{bermar,hentei-book}.
See \cite{bc,hht,holten,bhattacharya} for the alternative ``holomorphic representation.''

We look for a path integral representation of 

\bear
S^{x'x}[A] &=&
=
\big \langle x' \big|
{\bigl[m-i\slashed{D} \bigr] }^{-1}
\big| x \big\rangle
=
\big \langle x'\big|
{\bigl[m-i\s \partial + e\s A\bigr] 
}^{-1}
\big| x \big\rangle
=
\big\langle x'\big| {\bigl[m+\s p + e\s A\bigr] }^{-1} \big| x\big \rangle\, .
\nonumber\\
\label{defSbg}
\ear
We start with using the Gordon identity

\bear
{\slashed{D}}^2 =  -D_{\mu}D^{\mu} +{i\over 2}\, e \gamma^{\mu}\gamma^{\nu} F_{\mu\nu}\,,
\label{idgordon}
\ear
to rewrite

\bear
{\bigl[m-i\s{D} \bigr] }^{-1}
 &=& 
\bigl[m+i\s D\bigr] \bigl[m+i\s D\bigr]^{-1} {\bigl[m-i\s{D} \bigr] }^{-1}
\nonumber\\
&=&
\Bigl[m+i\s D\Bigr] \Bigl[m^2- D_{\mu}D^{\mu}
+{i\over 2}\, e \gamma^{\mu}\gamma^{\nu} F_{\mu\nu}\Bigr]^{-1} \, .
\nonumber\\
\label{diracrearrangeA}
\ear
This brings us to the formulas defining the second-order representation that we already quoted in the introduction,
\eqref{StoK} and \eqref{defK}. 
The kernel $K^{x'x}[A]$ is formally identical with the propagator for a scalar particle in the background containing the gauge field $A$ and
the matrix-valued potential $V = {i\over 2}\, e\gamma^{\mu}\gamma^{\nu}F_{\mu\nu}$. It is thus straightforward to obtain the following path integral
representation for it (see, e.g., \cite{41}) 

\bear
K^{x'x}[A] &=&
\int_0^{\infty}
dT\,
\e^{-m^2T}
\int_{x(0)=x}^{x(T)=x'}
Dx\,
{\cal P}
\e^{-\int_0^T d\tau\bigl(
\kinb
+ie\,\dot x\cdot A + 
{i\over 2}\, e\gamma^{\mu}\gamma^{\nu}F_{\mu\nu}
\bigr)}\, ,
\label{8SA1}
\ear\no
which generalises \eqref{Gammaspin} to the open-line case. 
We now wish to remove the path-ordering $\cal P$. 
This requires an identity like the one we used for the closed-loop
case, \eqref{idspinfactor}, but without taking the trace. 
As we show in appendix \ref{app-symb}, this identity is

\bear
{\cal P}
e^{-\int_0^T d\tau
{i\over 2}\, eF_{\mu\nu}\gamma^{\mu}\gamma^{\nu}
}
=
2^{-\frac{D}{2}}
{\rm symb}^{-1}
\int_{A}D\psi
\, \e^
{-\int_0^Td\tau\,
\bigl[\half\psi_{\mu}\dot\psi^{\mu}-ieF_{\mu\nu}(\psi+\eta)^{\mu}(\psi+\eta)^{\nu}\bigl]
}\, .
\label{idsfgen}
\ear
Here the symbol map, symb, is defined by 

\bear
{\rm symb} 
\bigl(\gamma^{\alpha_1\alpha_2\cdots\alpha_n}\bigr) \equiv 
(-i\sqrt{2})^n
\eta^{\alpha_1}\eta^{\alpha_2}\ldots\eta^{\alpha_n}\, ,
\label{defsymb}
\ear
%
where $\gamma^{\alpha\beta\cdots\rho}$ denotes the totally antisymmetrised product of gamma matrices:

\bear
\gamma^{\alpha_1\alpha_2\cdots \alpha_n} \equiv \frac{1}{n!}\sum_{\pi\in S_n} 
{\rm sign}(\pi) \gamma^{\alpha_{\pi(1)}}\gamma^{\alpha_{\pi(2)}} \cdots \gamma^{\alpha_{\pi(n)}} \, .
\label{Defantisymm}
\ear
Note that in $D$ dimensions the right-hand of \eqref{defsymb} side will vanish for more than $D$ factors by the Grassmann
property of the $\eta^{\alpha}$. 

Putting the pieces together, we arrive at our final path integral representation
for the kernel as given in the introduction, equation \eqref{Kfin}.  
Together with \eqref{StoK} it is a suitable starting point for calculating the 
fermionic propagator $S^{x'x}[A]$ in the string-inspired formalism. 

\section{Master formula for the \boldmath{$N$} - photon kernel in \boldmath{$x$} - space}
\label{Kx}

Choosing $A(x)$ as a sum of $N$ plane waves with polarisation vectors $\varepsilon_{i}^{\mu}$ and wave vectors $k_i^{\mu}$ as in (\ref{Kfin}), and
keeping only the terms containing each polarisation vector linearly, 
we get the ``$N$-photon dressed'' version of the kernel $K^{x'x}$:

\bear
K^{x'x}_N(k_1,\varepsilon_1;\ldots ; k_N,\varepsilon_N) &=&
(-ie)^N 
2^{-\frac{D}{2}}
\int_0^{\infty}
dT\,
\e^{-m^2T}
\e^{-\fourth \frac{(x-x')^2}{T}}
\int_{q(0)=0}^{q(T)=0}
Dq\,
\e^{-\int_0^T d\tau
\frac{\dot q^2}{4}}
\nonumber\\
&& \times 
{\rm symb}^{-1}
\int_{\psi(0)+\psi(T)=0
\hspace{-30pt}}D\psi
\, \e^
{-\int_0^Td\tau\,
\half\psi\dot\psi
}\,
 V^{x'x}_{\eta}[k_1,\varepsilon_1]\cdots  V^{x'x}_{\eta}[k_N,\varepsilon_N]\,.
 \nonumber\\
\label{Ndressed}
\ear\no
Here $ V^{x'x}_{\eta}[k,\varepsilon]$ is the photon vertex operator for the open line, which now reads 

\bear
 V^{x'x}_{\eta}[k,\varepsilon] 
=  \int_0^Td\tau \biggl\lbrack \varepsilon\cdot \Bigl( \frac{x'-x}{T} + \dot q\Bigr) 
+ 2i\varepsilon\cdot \bigl(\psi+ \eta\bigr) k\cdot \bigl(\psi+ \eta\bigr)
\biggr\rbrack \e^{ik\cdot \bigl( x+ (x'-x)\frac{\tau}{T} + q(\tau)\bigr)} \, .
 \nonumber\\
\label{vertexqpsi}
\ear
We could now do the double path integral as it stands, using the Green functions $G_B$ and $G_F$ and standard Gaussian combinatorics.
However, if we aim at a closed-form expression valid for any $N$, it will be necessary to find a suitable extension of the exponentiation formula \eqref{vertopexp} to the
fermionic case. As we explained already in the introduction for the closed-loop case, an elegant way to achieve this is though the introduction of $N=1$
worldline superspace, as motivated by the underlying worldline supersymmetry \eqref{susy}. 
Thus, introducing the worldline superfield 
\begin{eqnarray}
Q^{\mu}(\tau) \equiv q^{\mu} (\tau) 
+ \sqrt 2\,\theta\psi^{\mu}(\tau) \, ,
\label{defQ}
\end{eqnarray}
and using the superspace conventions introduced in the introduction,
we can rewrite the vertex operator \eqref{vertexqpsi} in the form

\bear
 V^{x'x}_{\eta}[k,\varepsilon] 
=  \int_0^Td\tau \int d\theta \,
\varepsilon\cdot \biggl\lbrack -\theta \, \frac{x'-x}{T} + \sqrt{2} \eta + DQ \biggr\rbrack 
\e^{ik\cdot \bigl[ x+ (x'-x)\frac{\tau}{T} + \sqrt{2}\theta \eta + Q(\tau)\bigr]}\, .
 \nonumber\\
\label{vertexsuperline}
\ear
Recall that for the time being we must also treat the polarisation vectors $\varepsilon_i$ as Grassmann variables.
After the usual formal exponentiation of the prefactor, we obtain the required purely exponential form
of the vertex operator:

\bear
V^{x'x}_{\eta}[k,\varepsilon]  =  \int_0^Td\tau \int d\theta 
\e^{ik\cdot x + \frac{x'-x}{T}(\theta \varepsilon+ i \tau k) - \sqrt{2}\eta\cdot (\varepsilon+ i \theta k)
+ \varepsilon \cdot DQ + ik\cdot Q}
\Big\vert_{\varepsilon}\, .
\ear
Thus the path integral is ready for evaluation by completion of the square, which yields 
the following Bern-Kosower type master formula for the $N$-photon kernel in $x$-space:

\bear
K^{x'x}_N(k_1,\varepsilon_1;\ldots ; k_N,\varepsilon_N) &=&
(-ie)^N 
{\rm symb}^{-1}
\int_0^{\infty}
\frac{dT}{(4\pi T)^{\frac{D}{2}}}
\e^{-m^2T}
\e^{-\fourth \frac{(x-x')^2}{T}}
 \int_0^Td\tau_1   \cdots  \int d\theta_N
 \nonumber\\
&& \hspace{-110pt} \times 
\e^{\sum_{i=1}^N \bigl[ ik_i\cdot x + \frac{x'-x}{T}(\theta_i \varepsilon_i + i \tau_i k_i) - \sqrt{2}\eta\cdot (\varepsilon_i+ i \theta_i k_i)\bigr]
+ \sum_{i,j=1}^N \bigl[\hat \Delta_{ij}k_i\cdot k_j +2iD_i\hat \Delta_{ij}\varepsilon_i\cdot k_j 
+ D_iD_j \hat \Delta_{ij} \varepsilon_i\cdot\varepsilon_j\bigr] }
\multilin\, .
 \nonumber\\
\label{Ndressedmasterx}
\ear\no
Here $\hat\Delta$ is now the super worldline Green's function appropriate for the combination of
Dirichlet and antiperiodic boundary conditions at hand:
\bear
 \hat \Delta(\tau,\theta;\tau',\theta') \equiv \Delta (\tau,\tau') + \half \theta \theta' G_F(\tau,\tau')\, .
 \ear
Let us also write explicitly the derivatives of this Green's function that appear in the master formulas:

\bear
D_i\hat\Delta_{ij} &=& \half\theta_j G_{Fij} - \theta_i \ddel_{ij}\, ; \nonumber\\
D_j\hat\Delta_{ij} &=& -\half\theta_i G_{Fij} - \theta_j \deld_{ij}\, ; \nonumber\\
D_iD_j\hat\Delta_{ij} &=& - \half G_{Fij} + \theta_i \theta_j \ddeld_{ij} \nonumber\\
\label{Dformulas}
\ear
(no summation convention).

\section{Master formula for the \boldmath{$N$} - photon kernel in momentum space}
\label{Kp}

In the following we give the momentum space version of the master formula derived above. We eventually specialise to $D = 4$ and work out the explicit form of the kernel for some simple cases.

\subsection{The master formula}

We begin by Fourier transforming the master formula (\ref{Ndressedmasterx}) to momentum space,
\bear
K^{p'p}_N(k_1,\varepsilon_1;\ldots ; k_N,\varepsilon_N) &=&
\int d^Dx\int d^Dx'\, {\rm e}^{ip\cdot x+ip'\cdot x'}\,
K^{x'x}_N(k_1,\varepsilon_1;\ldots ; k_N,\varepsilon_N) \,.
\label{Fourier}
\ear
After a change of variables from $x$, $x'$ to $x_{\pm}$ as in \eqref{cov}, 
the $x_+$ - integral produces the global momentum conservation factor $(2\pi)^D \delta^D(p+p'+\sum_{i=1}^N k_i)$, which we omit in the
following.  The $x_-$ - integral can, using momentum conservation, be written as

\bear
\int d^Dx_- \e^{-\fourth \frac{x_-^2}{T}+ \bigl[ ip' +\frac{1}{T}\sum_{i=1}^N(\theta_i \varepsilon_i + i \tau_i k_i)\bigr]\cdot x_- }
= 
(4\pi T)^{\frac{D}{2}}\e^{T \bigl[ ip' +\frac{1}{T}\sum_{i=1}^N(\theta_i \varepsilon_i + i \tau_i k_i)\bigr]^2} \, .
\label{dointxm}
\ear
This brings us to

\bear
K^{p'p}_N(k_1,\varepsilon_1;\ldots ; k_N,\varepsilon_N) &=&
(-ie)^N 
{\rm symb}^{-1}
\int_0^{\infty}
dT
\e^{-m^2T}
 \int_0^Td\tau_1   \cdots  \int d\theta_N\, \e^{\rm Exp}\multilin \, ,\nonumber\\
 \label{Ndressedmasterp}
 \ear
 where
 
 \bear
  {\rm Exp} &=& T \bigl[ ip' +\frac{1}{T}\sum_{i=1}^N(\theta_i \varepsilon_i + i \tau_i k_i)\bigr]^2- \sum_{i=1}^N \sqrt{2}\eta\cdot (\varepsilon_i+ i \theta_i k_i)
  \nonumber\\ && 
+ \sum_{i,j=1}^N \bigl[\hat \Delta_{ij}k_i\cdot k_j +2iD_i\hat \Delta_{ij}\varepsilon_i\cdot k_j 
+ D_iD_j \hat \Delta_{ij} \varepsilon_i\cdot\varepsilon_j\bigr] \,.
 \nonumber\\
\label{defExp}
\ear\no
Using (\ref{Dformulas}),(\ref{derDelta}) and momentum conservation, 
this can be written explicitly as (in the following we often abbreviate ${\rm sign}(\tau_i-\tau_j)$ by $\sigma_{ij}$ and $\delta\left(\tau_{i} - \tau_{j}\right)$ by $\delta_{ij}$)
%

\bear
{\rm Exp}
&=&- p'^2 T - \sum_{i=1}^N \sqrt{2}{\eta}\cdot (\varepsilon_i+ i \theta_i k_i) + \half\sum_{i,j=1}^N  \theta_i\theta_j \sigma_{ij} k_i\cdot k_j
\nonumber\\
&&
+
\sum_{i=1}^N (i\theta_i\varepsilon_i - \tau_i k_i)
\cdot\bigl(p'-p-\sum_{j=1}^N\sigma_{ij}k_j\bigr)
-i\sum_{i,j=1}^N \sigma_{ij}\varepsilon_i\cdot k_j\theta_j
 \nonumber\\ && - \half \sum_{i,j=1}^N \bigl(\sigma_{ij} + 2 \theta_i\theta_j\delta_{ij}\bigr)\varepsilon_i\cdot\varepsilon_j 
\, .
\label{exponentfin}
\ear
This appears to be the most useful form of writing the exponent of the momentum-space master formula. 
Nevertheless, let us mention in passing that there is also a suggestive form of the exponent that generalises \eqref{master-dss}. 
There we found that, in the scalar case, with the additional definitions \eqref{defKextension}
the exponent can be rewritten purely in terms of the functions $\vert \tau_i -\tau_j \vert$,
${\rm sign}(\tau_i-\tau_j)$ and $\delta(\tau_i-\tau_j)$, that is, in terms of the Green's function
for the second derivative on the line
\bear
g(\tau,\tau') \equiv \half \vert \tau -\tau' \vert\, ,
\label{defDeltaline}
\ear
and its derivatives. 
Worldline supersymmetry then leads one to suspect that, in spinor QED, 
a similar rewriting should be possible in terms of the supersymmetric generalisation of this
Green's function, and its super-derivatives. This Green function can be given in terms of the super-distance on the line,
$\vert \tau -\tau' \vert  + \theta\theta'{\rm sign}(\tau - \tau')$, as (see, e.g., \cite{polyakovbook, MeContact}):

\bear
\jhat g(\tau,\theta;\tau',\theta') &\equiv&\half (\vert \tau -\tau' \vert  + \theta\theta' {\rm sign}(\tau - \tau') ) \,
\label{defhatDeltaline}
\ear
so that
\bear
D_i\,\jhat g_{ij} &=& - \half (\theta_i-\theta_j)\sigma_{ij} \nonumber\\
D_iD_j\,\jhat g_{ij} &=& -\half (\sigma_{ij} + 2\theta_i\theta_j\delta_{ij})\, .\nonumber\\
\label{derhatDeltaline}
\ear
And indeed, further defining $\theta_0=\theta_{N+1}\equiv 0$ we can rewrite the kernel in the following, more compact way:
\bear
K^{p'p}_N(k_1,\varepsilon_1;\ldots ; k_N,\varepsilon_N) &=&
(-ie)^N 
{\rm symb}^{-1}
\int_0^{\infty}
dT
\e^{-m^2T}
 \int_0^Td\tau_1   \cdots  \int d\theta_N
 \nonumber\\
&& \hspace{-100pt} \times 
\e^{ - \sqrt{2}\eta \cdot \sum_{i=1}^N (\varepsilon_i+ i \theta k_i)
+\sum_{i,j=0}^{N+1} \bigl[\jhat g_{ij}k_i\cdot k_j +2iD_i\jhat g_{ij}\varepsilon_i\cdot k_j  + D_iD_j \jhat g_{ij} \varepsilon_i\cdot\varepsilon_j\bigr] }
\Big\vert_{\varepsilon_1\cdots \varepsilon_N}\, .
\label{Ndressedmasterpfin}
\ear\no
\subsection{The master formula for $D=4$}
So far everything we have done is valid in any even dimension. From now on we specialise to the four-dimensional case, which will allow us to process the
master formula further. The right-hand side of the symbol identity \eqref{defsymb} then can have at most four factors. Since moreover the kernel $K_N^{p'p}$ 
is even in the $\eta^{\alpha}$s (which is clear already from  the definition of the kernel in $x$ - space, \eqref{defK}, but is also easy to check
from  \eqref{Ndressedmasterp}), the symbol map will appear now only with zero, two or four $\eta^{\alpha}$s. Thus all we shall ever need is
\bear
{\rm symb}^{-1}(1) &=& \Eins \,;\nonumber\\
{\rm symb}^{-1}(\eta^{\alpha_1}\eta^{\alpha_2}) &=& - \frac{1}{4} \lbrack \gamma^{\alpha_1},\gamma^{\alpha_2}\rbrack \, ;\nonumber\\
{\rm symb}^{-1}(\eta^{\alpha_1}\eta^{\alpha_2}\eta^{\alpha_3}\eta^{\alpha_4}) &=& 
\frac{1}{96} \sum_{\pi\in S_4}{\rm sign}(\pi)\gamma^{\alpha_{\pi(1)}}\gamma^{\alpha_{\pi(2)}}\gamma^{\alpha_{\pi(3)}}\gamma^{\alpha_{\pi(4)}}= - \frac{i}{4}\varepsilon^{\alpha_1\alpha_2\alpha_3\alpha_4}\gamma_5\, .
\nonumber\\
\label{evalsymb}
\ear 
Expanding out the master formula \eqref{Ndressedmasterp}, \eqref{defExp} in powers of $\eta$, and using \eqref{evalsymb}, we can write
\bear
	K_N^{p^{\prime}p} &=&  (-ie)^N \frac{\mathfrak{K}_N^{p^{\prime}p}}{\left(p^{2} + m^{2}\right)\left(p^{\prime 2} + m^{2}\right)}\, ; \nonumber\\
	\mathfrak{K}_N^{p^{\prime}p} &=& 
	 A_N\bone + B_{N\ab}\sigma^{\alpha\beta} -i C_N\gamma_{5} \, ,
	\nonumber\\
\label{NdressedmasterpABC}
\ear\no
where $\sigma^{\alpha\beta}=\frac12 [\gamma^\alpha,\gamma^\beta]$ and
\bear
A_N &=&(p^2+m^2)(p'^2+m^2)
\int_0^{\infty}dT\e^{-m^2T} \int_0^Td\tau_1   \cdots  \int_0^T d\tau_N
 \e^{{\rm Exp}(\eta = 0)}\Big\vert_{\theta_N\cdots\theta_1\varepsilon_1\cdots \varepsilon_N }\, ;\nonumber\\
B_N^{\ab} &=& (p^2+m^2)(p'^2+m^2)
\int_0^{\infty}dT\e^{-m^2T} \int_0^Td\tau_1   \cdots  \int_0^T d\tau_N 
\nonumber\\ && \times
\half \sum_{i,j=1}^N(\veps_i+i\theta_ik_i)^{\alpha}(\veps_j+i\theta_jk_j)^{\beta}
\e^{{\rm Exp}(\eta = 0)}\Big\vert_{\theta_N\cdots\theta_1\varepsilon_1\cdots \varepsilon_N } \, ;\nonumber\\
C_N&=& (p^2+m^2)(p'^2+m^2)
\int_0^{\infty}dT\e^{-m^2T} \int_0^Td\tau_1   \cdots  \int_0^T d\tau_N
\nonumber\\ && \times
\frac{1}{4!} \sum_{i,j,k,l=1}^N\varepsilon(\veps_i+i\theta_ik_i,\veps_j+i\theta_jk_j,\veps_k+i\theta_kk_k,\veps_l+i\theta_lk_l)
\e^{{\rm Exp}(\eta = 0)}\Big\vert_{\theta_N\cdots\theta_1\varepsilon_1\cdots \varepsilon_N }.\nonumber\\
\label{defANBNCN}
\ear
Here we use the notation
$\varepsilon(a,b,c,d) \equiv \varepsilon^{\alpha\beta\gamma\delta}a_{\alpha}b_{\beta}c_{\gamma}d_{\delta}$.
The factors $(p^2+m^2)(p'^2+m^2)$ have been introduced for later convenience. The coefficient matrix $B_N^{\ab}$ will be taken to be antisymmetric.

We note that, comparing \eqref{defK} and \eqref{NdressedmasterpABC}, it is clear that the contribution to $\mathfrak{K}_N^{p^{\prime}p}$
involving $A_N$ has a part that by itself just gives, after dropping the unit matrix, the (truncated) dressed propagator in scalar QED. 
Thus we will denote this contribution by $A_N^{\rm scal}$, and write $A_N=A_N^{\rm scal}+A_N^{\psi}$. 

\subsection{Explicit form of the kernel for $D=4$ and $N = 0,1,2$}

Let us work out here the explicit form of the kernel for $N=0,1,2$, as illustrative examples and since these results will be needed for our
calculations below in any case. Here we use \eqref{Ndressedmasterp} and \eqref{defExp} rather than \eqref{Ndressedmasterpfin}.
The algebra is simple, starting with the expansion of the exponent and the truncation to the terms that are linear in all $\theta_i$ and
$\varepsilon_i$, only it should be kept in mind that all Grassmann variables (including the $d\theta_i$) anticommute with each other,
and that, to determine the absolute sign of the kernel, it is necessary to anticommute all the polarisation vectors to the left (or the right)
of all other Grassmann variables, and into the standard ordering $\varepsilon_1\cdots \varepsilon_N$.
Since we are computing the equivalent of tree-level diagrams in momentum space, it is furthermore clear
a priori that nontrivial or divergent parameter integrals cannot arise. 
In the following we will also set the electron charge $e=1$. 

For $K_0^{p'p}$ we find simply

\bear
K_0^{p'p} = {\rm symb}^{-1}\int_0^{\infty}dT\,\e^{-m^2T - p'^2 T} = \frac{\Eins}{p'^2+m^2} = \frac{\Eins}{p^2+m^2} \, ,
\label{K0}
\ear
which coincides with the scalar propagator of the second order formalism shown
in Figure~\ref{fig:secondorderrules}. 

For $N =1$ we find

\bear
K_1^{p'p} &=& (-i)\, {\rm symb}^{-1}\int_0^{\infty}dT\,\e^{-(m^2+p'^2)T}\int_0^Td\tau \int d\theta
\nonumber\\
&&\times   {\rm exp}\biggl\lbrace - \sqrt{2}\eta\cdot(\varepsilon+i\theta k) + (p'-p)\cdot (i\theta\varepsilon - \tau k)\biggr\rbrace
\lin
\nonumber\\
&=&
 (-i)\, {\rm symb}^{-1} \Bigl( i(p'-p)\cdot\varepsilon + 2i \varepsilon\cdot\eta k\cdot\eta \Bigr)
 \int_0^{\infty}dT\,\e^{-(m^2+p'^2)T}\int_0^Td\tau \e^{-\tau k\cdot(p'-p)}
 \nonumber\\
 &=& \frac{(p'-p)\cdot \varepsilon\Eins + \frac{1}{2} (\s k \s \varepsilon - \s \varepsilon \s k)}{(p^2+m^2)(p'^2+m^2)}
 = \frac{\s \varepsilon {\s {p}} - {\s p}' \s\varepsilon}{(p^2+m^2)(p'^2+m^2)}
 = \frac{\s \varepsilon ({\s {p}}-m) - (\s p'-m) \s\varepsilon}{(p^2+m^2)(p'^2+m^2)} \, ,
 \nonumber\\
 \label{K1}
\ear
which we shall later relate to the electron-photon vertex. Finally the calculation for $N=2$ leads in the first place to the integral representation

\bear
K_2^{p'p} &=&  (-i)^2{\rm symb}^{-1}\int_0^{\infty}dT\,\e^{-(m^2+p'^2)T}\int_0^Td\tau_1d\tau_2 
\nonumber\\
&& \times
\biggl\lbrace
4\veps_1\cdot\eta \veps_2\cdot\eta k_1\cdot\eta k_2\cdot\eta 
+2 k_1\cdot k_2 \veps_1\cdot\eta \veps_2\cdot\eta\,\sigma_{12} 
+2 \veps_1\cdot\veps_2 k_1\cdot\eta k_2\cdot\eta \sigma_{12}
\nonumber\\
&& \,\,\,\,
-2\bigl\lbrack \veps_1\cdot\eta k_1\cdot\eta \veps_2\cdot(p'-p) + (1 \leftrightarrow 2) \bigr\rbrack
-2\bigl\lbrack\veps_1\cdot\eta (k_1+k_2)\cdot\eta\veps_2\cdot k_1\sigma_{12} + (1 \leftrightarrow 2) \bigr\rbrack
\nonumber\\
&&  \,\,\,\,
- \veps_1\cdot(p'-p+\sigma_{21}k_2)\veps_2\cdot(p'-p+\sigma_{12}k_1) 
- \veps_1\cdot k_2 \veps_2\cdot k_1
+ k_1\cdot k_2 \veps_1\cdot\veps_2
+2\veps_1\cdot\veps_2 \delta_{12}
\biggr\rbrace
\nonumber\\
&&\,\,\times
\,\e^{k_1\cdot k_2 \abs{\tau_1-\tau_2} - (p'-p)\cdot (\tau_1k_1+\tau_2 k_2)} \, .
\label{K2}
\ear
Note that in \eqref{K2}, as well as in the final line of \eqref{K1}, the polarisation vectors have 
turned back into ordinary vectors, leaving the vector $\eta$ as the only anticommuting quantity. 

For $K_2^{p'p}$, due to the presence of the $\sigma_{ij}$ factors in the integrand
performing the parameter integrals now requires a case distinction between $\tau_1 \geq \tau_2$ and $\tau_1 \leq \tau_2$. 
From our starting point \eqref{Ndressed} it is clear that these two sectors differ only by an interchange of the two photons,
so that it is sufficient to calculate the contribution of the first one. Special treatment is needed for the last term in braces in \eqref{K2}, involving 
$\delta_{12}$; it corresponds to the contribution of the seagull vertex, and has to be split between the two sectors. 
Thus we have to calculate two integrals:
\begin{align}
\int_0^{\infty}dT\,\e^{-(m^2+p'^2)T}\int_0^Td\tau_1\int_0^{\tau_1} &d\tau_2 \,\e^{k_1\cdot k_2 (\tau_1-\tau_2) - (p'-p)\cdot (\tau_1k_1+\tau_2 k_2)}\nonumber\\
&= \frac{1}{(m^2+p^2)[m^2+(p'+k_1)^2](m^2+p'^2)}\,,
 \label{int1}\\[2mm]
\int_0^{\infty}dT\,\e^{-(m^2+p'^2)T}\int_0^Td\tau_1\int_0^T & d\tau_2 \, \delta(\tau_1-\tau_2)\,\e^{k_1\cdot k_2 \abs{\tau_1-\tau_2} - (p'-p)\cdot (\tau_1k_1+\tau_2 k_2)}
\nonumber\\
&=  \frac{1}{(m^2+p^2)(m^2+p'^2)}\, .
 \label{int2}
\end{align}
we can write $K_2^{p'p}$ as

\bear
K_2^{p'p} &=& \frac{1}{(m^2+p^2)(m^2+p'^2)} 
\biggl\lbrace 
-2\veps_1\cdot\veps_2 
+ 
\biggl\lbrack \frac{1}{m^2+(p'+k_1)^2}
\Bigl( 
\veps_1\cdot(p'-p-k_2)\veps_2\cdot(p'-p+k_1) 
\nonumber\\ &&
+\veps_1\cdot k_2\veps_2\cdot k_1 
-k_1\cdot k_2 \veps_1\cdot\veps_2 
+\half \veps_1\cdot\veps_2 [\s k_1,\s k_2] + \half k_1\cdot k_2 [\s\veps_1,\s \veps_2] 
\nonumber\\
&&
+\half \veps_1\cdot k_2 [\s\veps_2,\s k_1] 
-\half [\s\veps_1,\s k_2]\veps_2\cdot k_1
-\half\veps_1\cdot(p'-p-k_{2})[\s\veps_2,\s k_2] 
-\half [\s\veps_1,\s k_1]\veps_2\cdot(p'-p+k_{1})
\nonumber\\
&&
+i\gamma_5 \varepsilon(\veps_1,\veps_2,k_1,k_2)
\Bigr)
+ (1\leftrightarrow 2)
\biggr\rbrack
\biggr\rbrace \, .
\label{K2fin}
\ear
In the decomposition \eqref{NdressedmasterpABC} this reads

\bear
A_2^{\rm scal} &=& 
2\veps_1\cdot\veps_2 
-
\biggl\lbrack \frac{1}{m^2+(p'+k_1)^2}
\veps_1\cdot(p'-p-k_2)\veps_2\cdot(p'-p+k_1) 
+ (1\leftrightarrow 2)
\biggr\rbrack
\nonumber\\
A_2^{\psi} &=& 
-
\biggl\lbrack \frac{1}{m^2+(p'+k_1)^2}
+ (1\leftrightarrow 2)
\biggr\rbrack
\frac{1}{2} {\rm tr}(f_1f_2)
\nonumber\\
B_2^{\ab} &=& 
\frac{1}{m^2+(p'+k_1)^2}
\Bigl( 
- \veps_1\cdot\veps_2 \, k_1^{\alpha}k_2^{\beta} 
- k_1\cdot k_2 \,\veps_1^{\alpha}\veps_2^{\beta} 
- \veps_1\cdot k_2 \,\veps_2^{\alpha}k_1^{\beta} 
+ \veps_2\cdot k_1 \,\veps_1^{\alpha}k_2^{\beta}
\nonumber\\ &&
\hspace{3.5em}+ \veps_1\cdot(p'-p-k_{2})\veps_2^{\alpha}k_2^{\beta} 
+ \veps_2\cdot(p'-p+k_{1}) \veps_1^{\alpha}k_1^{\beta}
\Bigr) 
+ (1\leftrightarrow 2)
\nonumber\\
C_{2}
&=& 
\Bigl(\frac{1}{m^2+(p'+k_1)^2}+\frac{1}{m^2+(p'+k_2)^2}\Bigr) \veps_{1}^{\alpha}\veps_{2}^{\beta}k_{1}^{\gamma}k_{2}^{\delta}
\varepsilon_{\alpha\beta\gamma\delta}\, .
 \nonumber\\
\nonumber\\
\label{A2B2C2}
\ear
To compare with the standard formalism, one can complete the antisymmetrised products of Dirac matrices to full products, to arrive at

\bear
K_2^{p'p} = \frac{1}{(m^2+p^2)(m^2+p'^2)} 
&&\!\!\biggl\lbrace 
 \frac{1}{m^2+(p'+k_1)^2}
 \Bigl\lbrack
- \s\varepsilon_1(\s p'+\s k_1 + m)\s\varepsilon_2 (\s p-m) \nonumber\\
 &&- (\s p' - m)  \s\varepsilon_1\s\varepsilon_2 (\s p-m)
 +  (\s p' - m)  \s\varepsilon_1(\s p'+\s k_1 - m)\s\varepsilon_2 
 \Bigr\rbrack\nonumber\\
&&+ (1\leftrightarrow 2)
\biggr\rbrace \, .
 \label{K2alt}
 \ear
 For checking the equivalence of \eqref{K2fin} and \eqref{K2alt}, note that the first equation decomposes $K_2^{p'p}$
 in terms of the standard basis of the Dirac representation of the Clifford algebra, given by the 16 matrices
 $\lbrace \Gamma^A \rbrace \equiv \lbrace \Eins, \gamma^{\mu} , \sigma^{\mu\nu}, \gamma^{\mu}\gamma_5,\gamma_5\rbrace$, of which only the even subalgebra appears here, however. 
 The coefficients of the decomposition $X= x_A \Gamma^A$ of an arbitrary $4\times 4$ matrix $X$ in this basis can be obtained using the trace:
 
 \bear
x_A = \fourth \tr (X\Gamma_A)\, ,
\ear
where $\Gamma_A$ denotes the inverse of $\Gamma^A$. In this way one finds, for arbitrary Lorentz vectors $a,b,c,d$, the identity

\bear
\s a \s b \s c \s d &=& (a\cdot b\, c\cdot d - a\cdot c\, b\cdot d + a\cdot d\, b\cdot c)\Eins
-i\varepsilon (a,b,c,d)\gamma_5 \nonumber\\
&& -\half\Bigl([\s a,\s b] c\cdot d - [\s a,\s c] b\cdot d +  [\s a,\s d] b\cdot c +  [\s b,\s c] a\cdot d - [\s b,\s d] a\cdot c +  [\s c,\s d] a\cdot b \Bigr)\, .
\nonumber\\
\ear
Using this formula it is straightforward to go from \eqref{K2alt} to \eqref{K2fin}.

\section{Spin-orbit decomposition of the \boldmath{$N$} - photon kernel}
\label{spinorbit}

The vertex operator \eqref{vertexqpsi} representing the coupling of the fermion line to a photon separates this interaction into two parts:
the first part in the square brackets on the right-hand side is the same as for the scalar case, and thus must represent the orbital degree of freedom
of the fermion, the second one implements the fermion spin and we refer to this as the spin interaction. This suggests that useful physical information should be contained in
a decomposition of the kernel $K_N$ in terms of the number of spin interactions:
\bear
K_N &=& \sum_{S=0}^N K_{NS}\, ,	
\label{decompso}
\ear
where $K_{NS}$ denotes the contribution to the kernel involving $S$ spin and $N$-$S$ orbital interactions. In particular,
$K_{N0}$ coincides (up to the unit matrix in spin-space) with the kernel for scalar QED. 

While this decomposition could be extracted from our various superfield master formulas above, here we find it more convenient to
return to the component version of the fermionic path integral, equation \eqref{Ndressed}, and to draw on known results for the closed-loop
case. Let us denote by $V_{\eta}$ the spin part of the integrand of the vertex operator \eqref{vertexqpsi}, omitting the
exponential factor:
\bear
V_{\eta}[k,\varepsilon] \equiv  2i\varepsilon\cdot \bigl(\psi+ \eta\bigr) k\cdot \bigl(\psi+ \eta\bigr)
=  - i \bigl(\psi+ \eta\bigr) \cdot f \cdot  \bigl(\psi+ \eta\bigr)
\label{defVspinpart}
\ear
(note that, in the component formalism, the polarisation vectors remain ordinary commuting vectors throughout). 

For $\eta =0$, it is known from the closed-loop case how to Wick-contract a product of any number of such objects
in closed form \cite{berkos,strassler2,41}. Namely, define a ``fermionic bi-cycle of length $n$'' by
\bear
G_F(i_1i_2\ldots i_n) \equiv G_{Fi_1i_2}G_{Fi_2i_3}\cdots G_{Fi_ni_1}Z_n(i_1i_2\ldots i_n) \qquad (n\geq 2)\,,
\label{deffermionicbicycle}
\ear
where $Z_n$ was defined in \eqref{defZn}.
Then the Wick contraction of $S$ factors of $V_{\eta = 0}$ can be written as 
\bear
\hspace{-2em}
&&W_{\eta = 0}\argS \equiv  i^S\Bigl\langle V_{\eta=0}[k_1,\veps_1] \cdots V_{\eta=0}[k_S,\veps_S] \Bigr\rangle
\nonumber\\
&& 
=    \sum_{\rm partitions}
(-1)^{cy} G_F(i_1i_2\ldots i_{n_1})G_F(i_{n_1+1}\ldots i_{n_1+n_2})\cdots G_F(i_{n_1+\ldots + n_{cy-1}+1}\ldots i_{n_1+\ldots + n_{cy}})\, .
 \nonumber\\
\label{defW0}
\ear
Here in the last line the sum runs over products of up to $S$ bi-cycles, $cy =1,\ldots,S$, $cy$ denoting the number of cycles and
$n_k$ the length of the cycle $k$, and over
all {\it inequivalent} possibilities to distribute the indices $1,\ldots,S$ among the arguments of the bi-cycles. 
Here two bi-cycles are considered equivalent if their arguments can be identified by cyclic rotation and/or inversion;
e.g., $G_F(1234)$ is equivalent to $G_F(2341)$, $G_F(4321)$ and $G_F(3214)$, but inequivalent to $G_F(1243)$ and $G_F(1324)$
(inequivalent cycles first appear at the four-point level). 
Products of cycles are considered equivalent if all of their factors are equivalent. 
For example, 
\bear
W_{\eta=0} (k_1,\veps_1) &=& 0\, ; \nonumber\\
W_{\eta=0} (k_1,\veps_1;k_2,\veps_2) &=& - G_F(12) \, ;  \nonumber\\
W_{\eta=0} (k_1,\veps_1;k_2,\veps_2;k_3,\veps_3) &=& - G_F(123)\, ;   \nonumber\\
W_{\eta=0} (k_1,\veps_1;k_2,\veps_2;k_3,\veps_3;k_4,\veps_4) &=& - G_F(1234) - G_F(1243) - G_F(1324) \nonumber\\
&& + G_F(12)G_F(34) +  G_F(13)G_F(24) +  G_F(14)G_F(23)\, . \nonumber\\
\nonumber\\
\label{Weamp}
\ear
For arbitrary $S$, a closed-form expression for $W_{\eta=0}$ can be given in terms of a Pfaffian
determinant:
\bear
W_{\eta=0}(k_1,\veps_1;\ldots; k_S,\veps_S) = 2^S   
\sqrt{{\rm det}\Bigl(G_{Fij} v_i\cdot v_j \Bigr)}\, ,
\ear
where $i=1,\ldots, 2S$ and $\lbrace v_i \rbrace$ is the joined set of all momentum and polarisation vectors (in any ordering). 

Since the transition to $\eta \ne 0$ amounts only to the shift $\psi(\tau) \to \psi(\tau) + \eta$, it can be simply implemented by
adding, to the cycle products of \eqref{defW0}, all possible terms where cycles get broken into chains by insertions of $\eta$s. 
Defining a ``fermionic bi-chain of length $n$'' by
\bear
G_F\vert i \vert &\equiv & \eta f_i \eta\, \, ; \nonumber\\
G_F\vert i_1i_2\ldots i_n\vert &\equiv& 2 G_{Fi_1i_2}G_{Fi_2i_3}\cdots G_{Fi_{(n-1)}i_n} \eta f_{i_1} f_{i_2} \cdots f_{i_n} \eta \qquad (n\geq 2)
\label{deffermionicbichain}
\ear
we can generalise \eqref{defW0} to
\bear
&& W_{\eta}\argS \equiv  i^S\Bigl\langle V_{\eta}[k_1,\veps_1] \cdots V_{\eta}[k_S,\veps_S] \Bigr\rangle
\nonumber\\
&& 
=    \sum_{\rm partitions}
(-1)^{cy} G_F(i_1i_2\ldots i_{m_1})G_F(i_{m_1+1}\ldots i_{m_1+m_2})\cdots G_F(i_{m_1+\ldots + m_{cy-1}+1}\ldots i_{m_1+\ldots + m_{cy}})
 \nonumber\\
 && \hspace{5em} \times G_F\vert i_{m_1+ \ldots + m_{cy} +1}\ldots i_{m_1 + \ldots + m_{cy} + n_1}\vert  
\cdots
G_F\vert i_{m_1+ \ldots + m_{cy} +n_1+ \ldots n_{ch-1}+1}\ldots i_S\vert  \, ,
\nonumber\\
\label{defWeta}
\ear
where now $cy$ denotes the number of cycles, $ch$ the number of chains. Again the sum runs over all inequivalent partitions, where
for the chains the only equivalence relation is inversion, $G_F\vert i_1i_2\ldots i_n\vert = G_F\vert i_n \ldots i_2i_1\vert$.
Note that the sign of a term still depends only on the number of \textit{cycles} it contains. For example,
\bear
W_{\eta} (k_1,\veps_1) &=& G_F\vert 1\vert = \eta f_1 \eta \, ;\nonumber\\
W_{\eta} (k_1,\veps_1;k_2,\veps_2) &=& 
 - G_F(12) + G_F\vert 12 \vert + G_F\vert 1\vert G_F\vert 2\vert\, ,
 \nonumber\\
 &=& - \frac{1}{2} G_{F12}G_{F21}\tr (f_1f_2)+
 2G_{F12}\eta f_1f_2\eta + \eta f_1 \eta\, \eta f_2 \eta \, ;
 \nonumber\\
W_{\eta} (k_1,\veps_1;k_2,\veps_2;k_3,\veps_3) &=& - G_F(123) 
- \bigl[G_F(12)G_F\vert 3\vert + 2\, {\rm cycl.\,perm.}\bigr]  
+ G_F\vert 123\vert + G_F\vert 231\vert + G_F\vert 312\vert 
\nonumber\\&&
+ \bigl[G_F\vert 12 \vert G_F \vert 3\vert + 2\, {\rm cycl.\,perm.}\bigr]
+ G_F\vert 1\vert G_F\vert 2\vert G_F\vert 3\vert \, ,
\nonumber\\
&=&
- G_{F12}G_{F23}G_{F31}\tr (f_1f_2f_3) 
 - \Bigl[\frac{1}{2} G_{F12}G_{F21}\tr (f_1f_2)\eta f_3\eta + 2\, {\rm cycl.\,perm.}\Bigr]
\nonumber\\&& 
+ 2\bigl[G_{F12}G_{F23}\,\eta f_1f_2f_3\eta 
+ G_{F23}G_{F31}\,\eta f_2f_3f_1\eta + 
G_{F31}G_{F12}\,\eta f_3f_1f_2\eta]
\nonumber\\
&& + 2\bigl[G_{F12}\,\eta f_1f_2\eta \, \eta f_3\eta + 2\, {\rm cycl.\,perm.}\bigr] +  \eta f_1 \eta\, \eta f_2 \eta \, \eta f_3 \eta\, . \nonumber \\
\label{Weampeta}
\ear
Here we must remember once more that no more than $D$ factors of $\eta$ can appear in a term. Thus in four dimensions
the last term appearing in $W_{\eta} (k_1,\veps_1;k_2,\veps_2;k_3,\veps_3)$ above can already be omitted, since it carries six factors
of $\eta$. 

We now combine these results for the spin part with \eqref{Ndressed} and the results of subsection \ref{derivescalarmaster}
 to arrive at the
following explicit representation of the spin-orbit decomposition:
\bear
K_{NS} &=& \sum_{\lbrace i_1i_2\ldots i_S\rbrace} K_{NS}^{\lbrace i_1i_2\ldots i_S\rbrace}\, , \nonumber\\
K_{NS}^{\lbrace i_1i_2\ldots i_S\rbrace }  &=& (-ie)^N(-i)^N \symbi\int_0^\infty dT\,{\rm e}^{-m^2T}\prod_{i=1}^N\int_0^Td\tau_i\,
W_{\eta} (k_{i_1},\veps_{i_1};\ldots;k_{i_S},\veps_{i_S}) 
\bar P_{NS}^{\lbrace i_1i_2\ldots i_S\rbrace} 
\nonumber\\&&
\qquad \times {\rm e}^{-Tb_0^2+\sum_{i,j=1}^N \Delta_{ij}k_i\cdot k_j}\, .
\label{KNSexpl}
\ear
In the above the sum 
runs over all choices of $S$ out of the $N$ variables, and the bosonic prefactor polynomial $\bar P_{NS}^{\lbrace i_1i_2\ldots i_S\rbrace}$ 
is now defined by (compare \eqref{linemaster},\eqref{expred},\eqref{defb0})
\bear
\hspace{-3em}{\rm e}^{-Tb^2+\sum_{i,j=1}^N\big[\Delta_{ij}k_i\cdot k_j-2i\ddel_{ij}\varepsilon_i\cdot k_j-\ddeld_{ij}\varepsilon_i\cdot \varepsilon_j\big]}
\big \vert_{\veps_{i_1}= \cdots = \veps_{i_S} =0}\big \vert_{\varepsilon_{i_{S+1}}\cdots\varepsilon_{i_N}}
&\equiv &
(-i)^{N-S} \bar P_{NS}^{\lbrace i_1i_2\ldots i_S\rbrace } 
{\rm e}^{-Tb_0^2+\sum_{i,j=1}^N \Delta_{ij}k_i\cdot k_j}\, .
\nonumber\\
\label{defPNS}
\ear
Here the notation on the left-hand side means that one first sets the polarisation vectors $\veps_{i_1},\ldots,\veps_{i_S}$ equal to zero,
and then selects all the terms linear in the surviving polarisation vectors. 
In particular, one has the extremal cases
\bear
\bar P^{\lbrace \rbrace}_{N0} &=& \bar P_N \, ,\nonumber\\
\bar P^{\lbrace 12\ldots N\rbrace}_{NN} &=& 1 \, .\nonumber\\
\ear
Thus keeping only the $S=0$ term we get the $N$-photon kernel for scalar QED.
 
As an example, we arrive at the following concise rewriting of the the two-photon kernel, which was previous given in equation ~\eqref{K2}: 
\begin{align}
K_2^{p'p} &=  {\rm symb}^{-1} \int_0^\infty dT\, \e^{-(m^2+p'^2)T}\int_0^T d\tau_1 d\tau_2\,  \e^{k_1\cdot k_2 |\tau_1-\tau_2|-(p'-p)\cdot (\tau_1k_1+ \tau_2 k_2)}
\nonumber\\
&\times \Bigl\{ \bar P_2 +    W_{\eta}(k_1,\veps_1)\bar P_{21}^{\lbrace 1\rbrace}+ W_{\eta}(k_2,\veps_2)\bar P_{21}^{\lbrace 2\rbrace}
+ W_{\eta} (k_1,\veps_1;k_2,\veps_2)
\Bigr\}~.
\end{align}
Here $\bar P_2$ was given in \eqref{barP2}, $W_{\eta}(k_2,\veps_2)$ in \eqref{Weampeta} and
\bear
\bar P_{21}^{\lbrace1\rbrace} &=& 2(\ddel_{21}\veps_2\cdot k_1 + \ddel_{22} \veps_2\cdot k_2) - 2\veps_2 \cdot b_0 \, ,\nonumber\\
\bar P_{21}^{\lbrace2\rbrace} &=& 2(\ddel_{11}\veps_1\cdot k_1 + \ddel_{12} \veps_1\cdot k_2) - 2 \veps_1 \cdot b_0\, . \nonumber\\
\ear
Alternatively, in \eqref{KNSexpl} we can replace $\bar P^{\lbrace i_1i_2\ldots i_S\rbrace}_{NS}$ by the corresponding partially integrated 
$\bar Q^{\lbrace i_1i_2\ldots i_S\rbrace}_{NS}$ (note that the IBP procedure will
not generate terms with derivatives acting on the $G_{Fij}$ factors coming from the spin part).

\section{The dressed electron propagator in  momentum space}
\label{S}
Here we finally complete the transition from the second order formalism back to the familiar first order formalism by transforming $K_{N}$ to the physical $N$-photon dressed propagator of the Dirac field. 
\subsection{From $K$  to $S$}
The main object of interest in this paper is the dressed electron propagator in momentum space. A straightforward Fourier transformation of 
the $x$-space formulas \eqref{defSbg}, \eqref{diracrearrangeA} yields \eqref{SN}, which we repeat here for convenience:
\bear
S_N^{p'p} [\varepsilon_1,k_1;\ldots;\varepsilon_N,k_N]  &=& ({\s p'}+m) K_N^{p'p}[\varepsilon_1,k_1;\ldots;\varepsilon_N,k_N] 
\nonumber\\ &&
- e\sum_{i=1}^N{\s \varepsilon_i} K_{(N-1)}^{p'+k_i,p}[\varepsilon_1,k_1;\ldots;\hat \varepsilon_i,\hat k_i; \ldots\varepsilon_N,k_N] \, .
\label{SNdirect}
\ear
Here in the first term on the right-hand side all the polarisation vectors come from the kernel $K$, while in the others one was taken from the photon field contained in the covariant derivative acting on $K$ in formula \eqref{StoK}. 

Here it must also be remarked that our derivation of this identity contained some arbitrariness: in the first line of \eqref{diracrearrangeA} we could have placed
the factor $\bigl[m+i\slash D\bigr]$ to the right of the others, rather than to the left. If we do this, instead of \eqref{SN} we get the ``reversed'' identity
\bear
	S_N^{p^{\prime}p} [\varepsilon_1,k_1;\ldots;\varepsilon_N,k_N] & = & K_N^{p'p}[\varepsilon_1,k_1;\ldots;\varepsilon_N,k_N] (-\ps+m)
	\nonumber\\ && 
- e\sum_{i=1}^N K_{(N-1)}^{p',p+k_{i}}[\varepsilon_1,k_1;\ldots;\hat \varepsilon_i,\hat k_i; \ldots\varepsilon_N,k_N] {\s \varepsilon_i}\, .
\label{SNreversed}
\ear
Whichever of the two representations we use of the untruncated propagator $S_N^{p'p}$, for most purposes it will be necessary to eventually introduce also the truncated (or amputated)
one, which we denote by $\jhat{S}_N^{p'p}$. With our conventions, the two are related by
\bear
\jhat S_N^{p'p} &\equiv& (-{\s p'}+m) S_N^{p'p} (\s p +m),
\label{6defSNhat}
\ear
which simply removes the propagators associated to the external electron legs with momenta $p$ and $p'$.
\subsection{The cases $N=0,1,2$}

We will now extend our study of the cases $N=0,1,2$ in $D=4$ from the kernel to the propagator. 
This has the double purpose of studying how the equivalence with the standard first-order Feynman rules
comes about, and preparing our applications below. 

We start with $N=0$, that is the free propagator. 
Combining \eqref{K0} with \eqref{SN} gives
\bear
S_0^{p'p} &=& ({\s p'}+m)\frac{\Eins}{p'^2+m^2}  = ({\s p'}+m) ({\s p'}+m)^{-1} (-{\s p'}+m)^{-1} = ({\s p}+m)^{-1} \, .
\label{S0}
\ear
For $N=1$, eq. \eqref{SN} gives, using \eqref{K0} and \eqref{K1}, as well as
momentum conservation, we find
\bear
S_1^{p'p} &=& ({\s p'}+m) K_1^{p'p}[\varepsilon,k] 
- \s \varepsilon K_0^{p'+k,p}
\nonumber\\
&=&
({\s p'}+m)  \frac{\s \varepsilon ({\s {p}}-m) - (\s p'-m) \s\varepsilon}{(p^2+m^2)(p'^2+m^2)}
- \frac{ \s \varepsilon}{p^2+m^2}
\nonumber\\
&=&
\frac{({\s p'}+m) \s \varepsilon ({\s {p}}-m)}{(p^2+m^2)(p'^2+m^2)}\, .
\label{S1expl}
\ear
It is this result that in fact motivates (\ref{6defSNhat}) from which we get
\bear
\jhat S_1^{p'p} &=& (-{\s p'}+m) \frac{({\s p'}+m) \s \varepsilon ({\s {p}}-m)}{(p^2+m^2)(p'^2+m^2)} (\s p+m) = - \s \varepsilon\, ,
\ear
and we have reproduced the Dirac vertex as expected. 

For $N=2$,  eq. \eqref{SN} yields, now using \eqref{K1} and \eqref{K2alt},
 \bear
S_2^{p'p} &=& ({\s p'}+m) K_2^{p'p}[\varepsilon_1,k_1;\varepsilon_2,k_2]
-\s \varepsilon_1 K_1^{p'+k_1,p}[\varepsilon_2,k_2]
-\s \varepsilon_2 K_1^{p'+k_2,p}[\varepsilon_1,k_1]
\nonumber\\
&=&
\frac{({\s p'}+m)}{(p^2+m^2)(p'^2+m^2)} 
\biggl\lbrace 
 \frac{1}{(p'+k_1)^2+m^2}
 \Bigl\lbrack
- \s\varepsilon_1(\s p'+\s k_1 + m)\s\varepsilon_2 (\s p-m) - (\s p' - m)  \s\varepsilon_1\s\varepsilon_2 (\s p-m)
 \nonumber\\
 && \hspace{185pt}
 +  (\s p' - m)  \s \varepsilon_1(\s p'+\s k_1 - m)\s\varepsilon_2 
 \Bigr\rbrack
+ (1\leftrightarrow 2)
\biggr\rbrace
\nonumber\\
&& - \biggl\lbrace \s\varepsilon_1 \frac{\s \varepsilon_2 ({\s {p}}-m) - (\s p'+k_1-m) \s\varepsilon_2}{[(p'+k_1)^2+m^2](p^2+m^2)}+ (1\leftrightarrow 2)
\biggr\rbrace\, .
\nonumber\\
\label{S2expl}
\ear
It is then easy to verify that

\bear
\jhat S_2^{p'p} &=& (-{\s p'}+m) S_2^{p'p} (\s p +m) =
 \s\varepsilon_{1} \frac{\ps' + \ks_{1} + m}{\left(p' + k_{1}\right)^{2} + m^{2}}\s \varepsilon_{2} + \s\varepsilon_{2} \frac{\ps' + \ks_{2} + m}{\left(p' + k_{2}\right)^{2} + m^{2}}\s\varepsilon_{1} \,.
\ear
This is indeed what we get in the standard formalism from the two corresponding Feynman diagrams in Figure~\ref{fig-Comp}.
\begin{figure}
	\centering
		\includegraphics[width=0.6\textwidth]{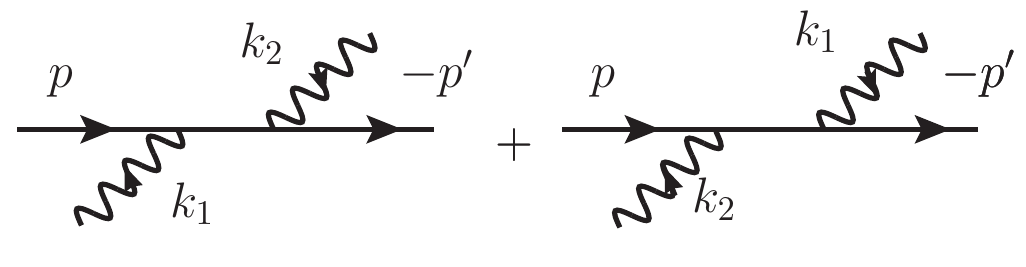}
		\caption{Feynman diagrams for the Compton scattering amplitude in the standard formalism.}
\label{fig-Comp}
\end{figure}

\section{The fermion self-energy}
\label{self energy}

Since the master formulae given in equations (\ref{SNdirect}, \ref{SNreversed}) hold off-shell, for the $N=2$ case they can, by sewing together the two photon legs, be used for the construction of the one-loop fermion self energy, indicated in Figure \ref{figSE}. 
We will carry out this calculation for an arbitrary dimension $D$ and gauge parameter $\xi$, and in
close analogy to the worldline calculation of the self-energy in scalar QED performed in \cite{102}. 
\begin{figure}[h]
	\centering
	\includegraphics[width=2.3in]{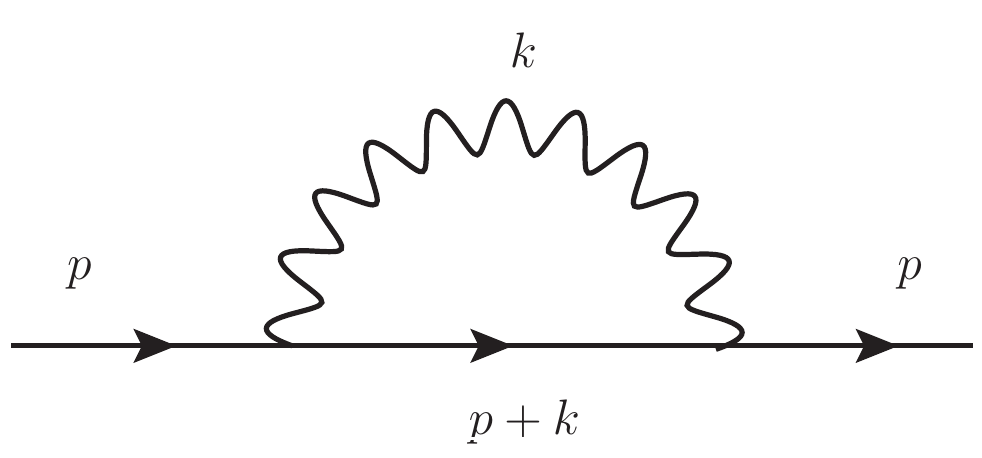}
	\caption{Electron self-energy diagram.}
	\label{figSE}
\end{figure}

\subsection{Construction of the self energy diagram by sewing} 

The dressed electron propagator in momentum space for $N=2$ is

\begin{align}
S_2^{p^\prime p} = \left( \slashed{p}^{\prime} + m \right) K_2^{p^\prime p} (k_1,\varepsilon_1; k_2, \varepsilon_2)
 - \slashed{\varepsilon}_1 K_1^{p^\prime +k_1, p} (k_2,\varepsilon_2)
- \slashed{\varepsilon}_2 K_1^{p^\prime +k_2, p} (k_1,\varepsilon_1) \label{dressed_propagator_2}\,.
\end{align}
We can immediately apply the decomposition (\ref{NdressedmasterpABC}) of $K_2$ with the explicit results for the coefficients in (\ref{A2B2C2}). Sewing consists of replacing 
$k_{1} = -k =-k_{2}$ which forces also $p' = -p$ and setting 
\begin{align}
\varepsilon_{1\mu}\varepsilon_{2\nu} \rightarrow \frac{\delta_{\mu\nu}}{k^{2}} - (1 - \xi)\frac{k_{\mu}k_{\nu}}{k^{4}}\, ,
\end{align}
which generates the photon propagator in an arbitrary covariant gauge. 
Following these substitutions, we then integrate over the loop momentum $k^{\mu}$.  

With these identifications it is easy to check that $C_{2}\big|_{-k_{1} = k = k_{2}} = 0$ and that
\begin{align}
	B_{2}^{\alpha \beta}\big|_{-k_{1}=k = k_{2}} \propto Dk^{\alpha}k^{\beta} + k^{2}\delta^{\alpha \beta} + 2 (1-\xi)k^{\alpha}k^{\beta}\, .
\end{align}
Multiplying this into the anti-symmetric matrix $\sigma^{\alpha \beta}$ gives a result that vanishes.
This leaves $A_{2}$ that is split into its scalar and spin parts that take the following form after sewing:

\begin{align}
	\hspace{-2em}A_{2}^{\textrm{scal}}\big|_{-k_{1} = k = k_{2}} &\rightarrow \frac{2}{k^{2}}\left[D - \frac{\left(2p + k\right)^{2}}{(p + k)^{2} + m^{2}} - \left(1 - \xi\right) \left(1 -\frac{ k\cdot (2p + k)\,\, (2p + k)\cdot k}{k^{2}\left((p + k)^{2} + m^{2}\right)} \right) \right]\, ,\nonumber\\
	\hspace{-2em}A_{2}^{\textrm{spin}}\big|_{-k_{1} = k = k_{2}} &\rightarrow -\frac{2 (D - 1)}{(p + k)^{2} + m^{2}}\,,
\end{align}
where we have taken advantage of the freedom to change the variable of integration $k \rightarrow -k$ to simplify the results. Note that
the spin contribution to $A$ is independent of the gauge parameter since the spin interaction is already written in terms of the field strength tensor, whilst the gauge dependent scalar piece is familiar from scalar QED -- see \cite{102}. Putting these together the contribution to the self energy from 
$A_2$ becomes ($d^{D}\bar{k} := \frac{d^{D}k}{(2\pi)^{D}}$)
\begin{equation}
	A_{2}^{\textrm{sew}} = \int \frac{d^{D}\bar{k}}{k^{2}} \left[D - \frac{(2p + k)^{2} + (D - 1)k^{2}}{(p + k)^{2} + m^{2}} + (1-\xi )\left( \frac{(k^{2} + 2 p \cdot k)^{2}}{k^{2}\left( (p + k)^{2} + m^{2}\right) } - 1 \right)\right].
\end{equation}
(we have dropped a factor of $2$ that is over- counted due to the permutation symmetry of external legs before the sewing takes place). 
Now, the very first and very last terms correspond to the diagrams with the seagull vertex and these vanish in dimensional regularisation. 
They can therefore be dropped so that (reinstating the electron charge)
\begin{align}
K^{p'p}_{(2,{\rm sew})}&=&e^2\int\frac{d^{D}\bar{k}}{k^{2}}\left[\frac{(2p + k)^{2} + (D - 1)k^{2}}{(m^2+p^2)^2[m^2+(p+k)^2]} -(1-\xi)\frac{(k^2+2p\cdot k)^2}{k^2(m^2+p^2)^2[m^2+(p+k)^2]}\right].
\label{K2sewn}
\end{align}
We must add to this the subleading terms.
Likewise using the $N = 1$ result, (\ref{K1}), applying the sewing procedure to $-\s\varepsilon_{1}K_{1}$ and $-\s\varepsilon_{2}K_{1}$ we find that each such term provides (here $j \neq i$)
\begin{equation}
 -\slash{\varepsilon}_{i} K_{(1, \textrm{sew})}^{p' + k_{i}, p}(k_{j}, \varepsilon_{j}) = \int \frac{d^{D}\bar{k}}{k^{2}}\left[\frac{(2 \ps - (D-2)\ks)}{(p^{2} + m^{2})((p + k)^{ 2} + m^{2})} - (1-\xi )\frac{\ks}{k^{2}} \frac{k^{2}+2p\cdot k}{(p^{2} + m^{2})((p + k)^{ 2} + m^{2})}\right].
\label{SubleadingIntK}
\end{equation}
After combining these terms with (\ref{K2sewn}) and using partial fraction decomposition, only five different integrals remain to be computed, and those
are already known from the scalar QED case \cite{102}:
\begin{align}
I_1&=\int \frac{d^Dq}{(2\pi)^D}\,\frac{1}{[m^2+(p+q)^2]}\,\,=\frac{(m^2)^{\frac{D}{2}-1}}{(4\pi)^{\frac{D}{2}}}\Gamma\Big(1-\frac{D}{2}\Big)\,;\non
I_2&=\int \frac{d^Dq}{(2\pi)^D}\frac{1}{q^2[m^2+(p+q)^2]}=-\frac{(m^2)^{\frac{D}{2}-2}}{(4\pi)^{\frac{D}{2}}}\Gamma\Big(1-\frac{D}{2}\Big)\,_2F_{1}\Big(2-\frac{D}{2},1;\frac{D}{2};-\frac{p^2}{m^2}\Big)\, ;\non
I_3^\mu&= \int \frac{d^Dq}{(2\pi)^D}\frac{q^\mu}{q^2[m^2+(p+q)^2]}=-\frac{p^{\mu}}{2p^{2}}\left[I_{1} + (p^{2} + m^{2})I_{2}\right]\,; \nonumber \\
J_1&=\int\frac{d^Dq}{(2\pi)^D}\frac{1}{q^4[m^2+(p+q)^2]}=\frac{(m^2)^{\frac{D}{2}-3}}{(4\pi)^{\frac{D}{2}}}\Gamma\Big(1-\frac{D}{2}\Big)\,_2F_1\Big(3-\frac{D}{2},2;\frac{D}{2};-\frac{p^2}{m^2}\Big)\, ;\non
J_2^\mu&=\int\frac{d^Dq}{(2\pi)^D}\frac{q^\mu}{q^4[m^2+(p+q)^2]}= -\frac{p^{\mu}}{2p^{2}}\left[I_{2} + (p^{2} + m^{2})J_{1}\right]\,.\nonumber \\
\end{align}
%
\no
In terms of these integrals, we can write the two contributions to $S_{(2,{\rm sew})}^{p'p}$ as 

\begin{align}
 ({\s p}^\prime +m)K^{p'p}_{(2,{\rm sew})}&=e^2\frac{{\s p}^\prime +m}{(m^2+p^2)^2}\Big[4p^2 I_2 + 4p\cdot I_3+DI_1+ (m^2+p^2)^2(\xi-1)J_1\Big]
\end{align}

and
\begin{equation}
	-\s \varepsilon_{(i, \rm sew)}K^{p'+k_j,p}_{(1, \rm sew)} 
	=\frac{e^2}{m^2+p^2}\Big[2\s pI_2-(D-2)\s I_3 -(\xi-1)(m^2+p^2)\s J_2\Big].
\end{equation}


Using the integration results above, we may write the contribution to the self energy in the following way:
%
\begin{align}
\hspace{-4.5em}	S^{p'p}_{(2,{\rm sew})} &=e^2\frac{({\s p}^\prime +m)}{(m^2+p^2)^2}\frac{(m^2)^{\frac{D}{2}-2}}{(4\pi)^{\frac{D}{2}}}\Gamma\Big(1-\frac{D}{2}\Big)\Bigg\{(D-2)m^2+2(m^2-p^2)\,_2F_1\Big(2-\frac{D}{2},1;\frac{D}{2};-\frac{p^2}{m^2}\Big)\non
\hspace{-4.5em}	&\hspace{5cm}+(\xi-1)\frac{(m^2+p^2)^2}{m^2}\,_2F_1\Big(3-\frac{D}{2},2;\frac{D}{2};-\frac{p^2}{m^2}\Big)\Bigg\}\non
\hspace{-4.5em}&-e^2\frac{{\s p}'}{2p^{2}(m^2+p^2)}\frac{(m^2)^{\frac{D}{2}-2}}{(4\pi)^{\frac{D}{2}}}\Gamma\Big(1-\frac{D}{2}\Big)\Bigg\{-\left[4p^2+(D-2)(m^2+p^2)\right]\,_2F_1\Big(2-\frac{D}{2},1;\frac{D}{2};-\frac{p^2}{m^2}\Big)+(D-2)m^2\non
\hspace{-4.5em}	&-(\xi-1)(m^2+p^2)\left[\,_2F_1\Big(2-\frac{D}{2},1;\frac{D}{2};-\frac{p^2}{m^2}\Big)-\frac{m^2+p^2}{m^{2}}\,_2F_1\Big(3-\frac{D}{2},2;\frac{D}{2};-\frac{p^2}{m^2}\Big)\right]\Bigg\}.
\label{SigmaIntegrated}
\end{align}
We are not quite done, however, as we should amputate the external fermions according to (\ref{6defSNhat})
\bear
\jhat{S}\,^{p'p}_{(2,\rm sew)}=(-\s p'+m)\,S^{p'p}_{(2,\rm sew)}\,(\s p+m)\, ;
\ear
doing this we can decompose the final result to 
(our notation follows \cite{Dav})

\begin{equation}
\jhat{S}\,^{p'p}_{2,\rm sew} = \alpha(p^2,D){\s p}^\prime +\beta(p^2,D)\bone \, \non
\end{equation}
where

\begin{align}
\alpha (p^2,D)&=\frac{e^2}{2p^2}\frac{(m^2)^{\frac{D}{2}-2}}{(4\pi)^{\frac{D}{2}}}\Gamma\Big(1-\frac{D}{2}\Big)(D-2)\bigg\{\,_2F_{1}\Big(2-\frac{D}{2},1;\frac{D}{2};-\frac{p^2}{m^2}\Big)\,(m^2-p^2)\Big[1+\frac{\xi-1}{D-2}\Big]-m^2\non
&\hspace{5cm}-\,_2F_{1}\Big(3-\frac{D}{2},2;\frac{D}{2};-\frac{p^2}{m^2}\Big)\frac{(m^2+p^2)^2}{m^2}\frac{\xi-1}{D-2}\bigg\}\, ,\non
\beta (p^2,D)&=\frac{e^2(m^2)^{\frac{D}{2}-2}\,m}{(4\pi)^{\frac{D}{2}}}\Gamma\Big(1-\frac{D}{2}\Big)(D+\xi-1)\,_2F_{1}\Big(2-\frac{D}{2},1;\frac{D}{2};-\frac{p^2}{m^2}\Big)\,.
\nonumber\\
\label{na}
\end{align}
Further simplification can be achieved by using the following identity for the hypergeometric function $_{2}F_{1}$ which we prove in appendix \ref{app-hypergeo}:

\begin{equation}
	_{2}F_{1}(a, 1, 2-a; -z)(1 - z)(1-2a) +{} _{2}F_{1}(a+1, 2, 2-a; -z)(1 + z)^{2} = 2(1-a)\, ,
	\label{idhypergeo}
\end{equation}
so that with $a = 2 - \frac{D}{2}$ and $z = \frac{p^{2}}{m^{2}}$ we get 

\begin{equation}
	{}_{2}F_{1}\Big(2 - \frac{D}{2}, 1, \frac{D}{2}; -\frac{p^{2}}{m^{2}}\Big)(m^{2} - p^{2})(D-3) 
	+ {}_{2}F_{1}\Big(3 - \frac{D}{2}, 2, \frac{D}{2}; -\frac{p^{2}}{m^{2}}\Big)\frac{(m^{2} + p^{2})^{2}}{m^2} = (D-2)m^{2}.
	\label{idF}
\end{equation}
Applying this identity to the coefficient function $\alpha(p^2,D)$ we obtain the simpler representation\footnote{As an aside, we note that 
the same identity can be used to simplify the expressions given in \cite{102} for the self energy and vertex in scalar QED. 
E.g. the scalar self energy can be rewritten (in our present notation)

\begin{align}
	& \frac{e^{2} (m^{2})^{\frac{D}{2} - 2} } {(4\pi)^{\frac{D}{2}}}\Gamma\left(1 - \frac{D}{2}\right)\left[m^{2} - 2(m^{2} - p^{2}) _{2}F_{1}\Big(2 - \frac{D}{2}, 1;\frac{D}{2}; -\frac{p^{2}}{m^{2}} \Big) + (1 - \xi)\frac{(p^{2} + m^{2})^{2}}{m^{2}} {}_{2}F_{1}\Big(3 - \frac{D}{2}, 2; \frac{D}{2}; -\frac{p^{2}}{m^{2}} \Big)\right] \nonumber  \\ 
	& = \frac{e^{2} (m^{2})^{\frac{D}{2} - 2} } {(4\pi)^{\frac{D}{2}}}\Gamma\left(1 - \frac{D}{2}\right) \left[m^{2}\left[1 + (1-\xi)(D-2)\right] - (m^{2} - p^{2}) _{2}F_{1}\Big(2 - \frac{D}{2}, 1;\frac{D}{2}; -\frac{p^{2}}{m^{2}} \Big) \left[2 + (1 - \xi)(D-3) \right] \right]\, .
\end{align}
}

\begin{eqnarray}
\alpha (p^2,D)
=\frac{e^2}{2p^2}\frac{(D-2)(m^2)^{\frac{D}{2}-2}}{(4\pi)^{\frac{D}{2}}}\Gamma\Big(1-\frac{D}{2}\Big) \xi \bigg\{(m^{2} - p^{2}) \,_2F_{1}\Big(2-\frac{D}{2},1;\frac{D}{2};-\frac{p^2}{m^2}\Big) - m^{2} \bigg\}\, .\non
\label{alphafinal}
\end{eqnarray}
In particular, it can now be seen that the coefficient function $\alpha(p^2,D)$ is absent for $\xi = 0$ (Landau gauge). 

Davydychev {\it et al.} have computed the self-energy in an arbitrary gauge and dimension for a non-Abelian $SU(N)$ theory \cite{Dav}. 
We find complete agreement with their results after putting their group parameter $C_F=1$ for the $U(1)$ symmetry of QED and transforming to Euclidean space 
(note that their gauge parameter, $\xi_{D}$, is related to ours by $\xi_{D} = 1 - \xi$).

\subsection{Special gauge choices}
Given that our treatment of the propagator naturally splits it up into the two terms that we have referred to as leading and subleading, we pause here to discuss a natural question that arises with respect to the gauge parameter, $\xi$, that we have so far left arbitrary. We will show that it is possible to choose $\xi$ such that either one of these pieces vanishes. 

Firstly we consider removing the subleading piece. This cannot be done at the level of the integrand in (\ref{SubleadingIntK}) so we instead consider the final two terms in (\ref{SigmaIntegrated}). Applying (\ref{idF}), one is led to the following value of $\xi$ that makes these two terms cancel, which we call $\xi_1(p^2,D)$:

\begin{eqnarray}
	\xi_1(p^2,D) = 1 +  \frac{[4p^{2} + (D-2)(p^{2} + m^{2})]{}_2F_{1}(2-\frac{D}{2}, 1; \frac{D}{2}; -\frac{p^{2}}{m^{2}}) - (D-2)m^{2} }{(D-2)m^{2} - {}_2F_{1}(2-\frac{D}{2}, 1; \frac{D}{2}; -\frac{p^{2}}{m^{2}}) [m^{2} + p^{2} + (D-3)(m^{2} - p^{2})]}.
	\nonumber\\
	\label{xisubleading}
\end{eqnarray}
However, this gauge parameter cannot be used in $D=4$, since the denominator becomes singular. In fact, in four dimensions the $\frac{1}{\epsilon}$ - pole of the subleading term 
is gauge independent, and proportional to the expression

\begin{equation}
	 \frac{e^{2}}{m^{2} + p^{2}}\left[(p^{2} - m^{2})\ps + 2m p^{2}\right]\, .
\end{equation}
The gauge parameter can, however, be used for QED in $D = 2$ dimensions, where it becomes
\bear
\xi_1(p^2,D) = -1 -2 (D-2) + \ldots
\label{xi12D}
\ear
Here we give also the linear term in the $\epsilon$ expansion since, due to the pole contained in the prefactor $\Gamma(1-\frac{D}{2})$ in 
\eqref{SigmaIntegrated}, it will have to be included if one wishes to remove the subleading term completely. 

For the leading contribution, the analysis is the same. We find that it vanishes for a gauge parameter $\xi_2(p^2,D)$,

\begin{equation}
\xi_2(p^2,D) = 1 + \frac{(D-2)m^{2} + 2(m^{2} - p^{2}){}_2F_{1}(2-\frac{D}{2}, 1; \frac{D}{2}; -\frac{p^{2}}{m^{2}})}{(D-3)(m^{2} - p^{2}){}_2F_{1}(2-\frac{D}{2}, 1; \frac{D}{2}; -\frac{p^{2}}{m^{2}}) - (D-2)m^{2}}\, .
\end{equation}
This time the gauge parameter does not become singular in four dimensions, and expanding around $D=4$ we find that the leading contribution 
to the propagator can be removed using

\begin{equation}
\xi_2(p^2,D) =  3 \frac{p^{2} - m^{2}}{m^{2} + p^{2}} + \frac{p^{2} - m^{2}}{m^{2} + p^{2}}\left[3\frac{m^{2}}{p^{2}}\log\left(1 + \frac{p^{2}}{m^{2}}\right) - 2\right] (D-4) + \ldots
\label{xi24D}
\end{equation}
Note that, in the massless limit, this becomes $\xi_2(p^2,D) \rightarrow 3 - 2(D - 4)$, whose constant term corresponds to Yennie-Fried gauge, $\xi = 3$. 
On the other hand, expanding around $D=2$ we find
\bear
\xi_2(p^2,D) =  -1  + \frac{3p^{2} -m^{2}}{m^{2} - p^{2}}(D-2)+\ldots
\label{xi22D}
\ear
Thus in $D=2$ both gauge parameters start with $\xi=-1$, so that here we can achieve more than in four dimensions: we can remove the pole of
the leading and subleading contribution simultaneously,  and the finite part of one or the other. This does not come unexpected, since it had been noted
already in \cite{94} that the $\xi = -1$ gauge in two dimensions has the property of removing the divergence of the one-loop fermion propagator 
(which in two dimensions is an IR one). More recently, this property has turned out to be extremely useful for multiloop calculations in the Schwinger model \cite{121}.
It will be interesting to see whether further simplification can be achieved by one of the generalisations \eqref{xi12D}, \eqref{xi22D}. 
 
Another open question is whether there exist similar choices of gauge that can remove various contributions at higher order, since it would be advantageous to have the option of removing the leading term, especially when considering amplitudes with a large number of photons attached to the line. This is because the leading contribution at order $N$, $({\s p}^\prime + m)K_{N}$ involves the $N$-photon kernel which is progressively more complicated than the $N$ subleading contributions of the form $\s{\varepsilon}_{i}K_{N-1}$. This is clear even in the results for $N = 1$ or $N = 2$ presented above in section \ref{S}. At higher order the simplifications gained by being able to discount $K_{N}$ could be substantial and may help to streamline various calculations. We leave this for examination in future work.

\section{Conclusions and Outlook}
\label{conc}

In this article we have presented a new -- and long overdue -- approach to the worldline path integral representation
of the open Dirac-fermion line dressed with $N$ photons. The formalism is designed to
extend to the open-line case the main calculational advantages of the well-established worldline 
formulation of the closed fermion loop, such as:

\begin{enumerate}
\item
Making possible the derivation of compact master formulas representing
whole classes of Feynman diagrams differing by the ordering of the photon
legs along a loop or line.
\item
Keeping a close analogy between scalar and spinor QED calculations, in particular
with respect to the simple dependence on the loop mass. 
\item
Minimising the effort in Dirac algebra manipulations through the use of the symbol map,
which effectively avoids long products of Dirac matrices by an early projection onto the 
Clifford basis.
\item
Allowing the generation of gauge-invariant structures by integration-by-part algorithms, rather than
the usual tedious analysis of the QED Ward identities.

\end{enumerate}
Our formalism is based on the second-order approach to spinor QED, which has been known
for decades as an alternative to the standard Dirac approach \cite{hostler,morgan} but rarely been 
considered as an alternative for state-of-the-art calculations (although in recent years 
it has been used as a starting point for the construction of non-standard abelian gauge theories
\cite{Napsuciale_1,Napsuciale_2,Mauro,Mariana_1,Mariana_2,JEspin}). It is also close in spirit to first-quantised string theory, and thus shares some of the
superior organisation of string amplitudes, particularly with respect to gauge invariance, permutation symmetry and worldline supersymmetry.  

In the present first part of this series of papers we have focused 
on the construction of a Bern-Kosower
type master formula for the fermion propagator dressed with $N$ photons, still off-shell and geared
towards the construction of multiloop amplitudes. We have given this formula in two versions, once
using worldline superfields and once via a spin-orbit decomposition that should contain additional
physical information. Both versions are amenable to numerical implementation. 
We have explicitly worked out the cases $N=0,1,2$, and demonstrated in detail how the
equivalence to the standard approach works. The $N=2$ result has further been used for
a recalculation of the one-loop fermion self energy for arbitrary dimension and arbitrary 
gauge parameter $\xi$. Cancellations for special values of $\xi$ have been found that look
promising for investigation at higher-loop order. 

The forthcoming second part will focus on on-shell amplitudes and cross sections involving open fermion lines, 
and in the third part we will add an external constant field (partial results of the third part have
already been published in \cite{113,ahmedwild-19}). 

In an independent publication we will use the formalism for an extension of the generalised
$2N$-point Landau-Khalatnikov-Fradkin transformation introduced in \cite{102} for scalar
QED, to the spinor QED case. 
Future additional articles will further be devoted to the application of the formalism to
multi-loop $g-2$ calculations, and to the derivation of Ball-Chiu form factors. 
Generalisations to the non-abelian case and to the inclusion of axial couplings are also 
under consideration. 

\acknowledgments

We would like to thank D.M. Gitman, D.G.C. McKeon and M. Reuter for discussions and useful correspondence. 
CS and JPE thank CONACYT for support through project Ciencias Basicas 2014 No. 242461. NA is grateful to IBS and CoReLS in South Korea where part of this work carried out. JPE thanks P. Cvitanovi\'{c} for useful conversations and further acknowledges financial support from U.M.S.N.H. through CiC project \#483224-2019. VMBG received support from PRODEP project 511-6/19-4990 for part of this work. 

\appendix
\section{Conventions}
\label{app-conv}
On the side of the worldline formalism, we work throughout in Euclidean space with metric $(+ + + \, +)$, and use
Dirac matrices fulfilling $\lbrace \gamma^{\mu},\gamma^{\nu}\rbrace = -2 \delta^{\mu\nu}$.
On the field theory side, we Wick rotate to Minkowski space with metric $\eta_{\mu\nu} = \textrm{diag}(- + + \, +)$, 
and use $\lbrace \gamma^{\mu},\gamma^{\nu}\rbrace = - 2 \eta^{\mu\nu}$. We further 
define $\varepsilon^{0123} = + 1$ and $\gamma_5 = i\gamma^0\gamma^1\gamma^2\gamma^3$. 
The fermion propagator becomes $-i/(\s p + m)$ and the first-order Dirac 
vertex $-ie\gamma^{\mu}$. 
The sign of the effective action corresponds to a tree-level term $-\frac{1}{4}F_{\mu\nu}F^{\mu\nu}$
in both Euclidean and Minkowskian spacetimes. 
The covariant derivative is $D_{\mu} = \partial_{\mu} + ieA_{\mu}$. 
These Minkowski space conventions coincide with the textbook of Srednicki \cite{srednicki-book}
except for the sign of the electric charge and 
that we use ingoing momenta in Feynman diagrams instead of outgoing ones. 
The Feynman rules for the second-order formalism have been given in the introduction, Figure \ref{fig:secondorderrules}.

\section{Intrinsic worldline approach to the electron propagator}
\label{app-intrinsic}

In this appendix, we rederive the path-integral representation of the electron propagator
in a more ``principled'' way, using the principles of quantum mechanics, gauge theory and (worldline) supersymmetry but no field-theory input. 

As is well-known, a spin 1/2 particle can be described in a manifestly covariant way by
a gauge model with one local supersymmetry on the worldline. For the massless case, the  phase space action 
depends on the particle space time coordinates $x^\mu$ joined by the real Grassmann variables $\psi^\mu$,
supersymmetric partners of the former that supply the degrees of freedom associated to spin.
In addition, there are Lagrange multipliers $e$ (the einbein) and $\chi$ (the gravitino),
with commuting and anti-commuting character, respectively, that gauge suitable first class constraints (they form the supergravity multiplet in  one dimension). Eventually,
their effect is to eliminate negative norm states from the physical spectrum, and make the particle model consistent with unitarity at the quantum level.

The action for the massless particle takes the form (given here in Minkowski space)
 
\bear
S \eqa \int d\tau 
\left (  
p_\mu \dot x^\mu + \frac{i}{2}  \psi_\mu \dot \psi^\mu  - e H - i \chi Q
\right )\, ,
\label{7.53}
\ear
where the first class constraints are given by

\bear
H= \frac12 p^2    \;, \qquad 
Q= p_\mu \psi^\mu \,,
\ear
that generate through Poisson brackets the $N=1$ susy algebra in one dimension

\bear
\{Q,Q\}= -2i H \;.
\ear
This algebra is computed by using the graded Poisson brackets of the phase space coordinates, 
$\{x^\mu, p_\nu \}=\delta^\mu_\nu$ and  $\{\psi^\mu,\psi_\nu\} = -i \delta^\mu_\nu $,
fixed by the symplectic term of the action.

The gauge transformations are generated on the phase space coordinates
$(x,p, \psi)$ through Poisson brackets with
$ V \equiv \zeta H + i\epsilon  Q$, where  $\zeta$ and $\epsilon $ are local parameters with appropriate 
Grassmann parity whilst on gauge fields the gauge transformations are obtained by using the structure constants of the constraint algebra and turn out to be

\begin{align}
\delta x^\mu = \zeta p^\mu + i\ep \psi^\mu  &\,;\qquad  \delta p_\mu= 0 \, ; \qquad \delta \psi^\mu = -\epsilon p^\mu \,;\\
\delta e &= \dot \zeta + 2 i \chi \ep \, ; \quad \delta \chi = \dot \ep  \ . 
\end{align}

Let us now study canonical quantisation to uncover the consequences of the constraints,
and see how the Dirac equation emerges.
Promoting the phase space variables to operators one finds the following (anti) commutation 
relations

\begin{equation}
[ \hat x^\mu , \hat p_\nu ] = i \delta^\mu_\nu \ ,
\quad \quad
\{ \hat \psi^\mu, \hat \psi^\nu \} = \eta^{\mu\nu} ,
\label{7.8a}
\end{equation}
while other graded commutators vanish.
The former relations are realised on the usual infinite dimensional Hilbert space of functions of the particle coordinates. 
The latter relations are seen to give rise to a Clifford algebra that may be identified 
with the algebra of the Dirac gamma matrices $\Gamma^\mu$, satisfying $\{ \Gamma^\mu, \Gamma^\nu\} = 2\eta^{\mu\nu}$ and as such they can be realised on the finite dimensional Hilbert space of spinors as

\begin{equation}
\hat \psi^\mu  \ \to \ \frac{1}{\sqrt{2}} \, \Gamma^\mu\, ,
\end{equation}
with dimension $2^{\left [\frac{D}{2}\right ]}$.
The full Hilbert space is the direct product of the two Hilbert spaces obtained above and is identified with the space of spinor fields.

The full information of the physical states, $\left| \Psi\right>$, resides in the constraints implemented \`a la Dirac. In particular, the constraint due to the
susy charge $\hat Q= \hat  p_\mu \hat \psi^\mu $ gives rise to the massless Dirac equations

\begin{equation}
\hat  p_\mu \hat \psi^\mu |\Psi \ra =0  \quad \to \quad  \Gamma^\mu \partial_\mu \Psi(x) =0.
\end{equation}
Likewise the constraint $\hat H| \Psi\ra =0 $ leads to the massless Klein Gordon equation  for all components of the spinor  $\Psi$,
and is automatically satisfied as a consequence of the algebra  $ \hat Q^ 2 = \hat H$. 
Thus, we recognise how a first quantised description of a spin 1/2 particle emerges from canonical quantisation of a constrained system.

To study the corresponding path integral quantisation it is useful to 
eliminate the momenta $p_\mu$ to obtain the action in configuration space

\begin{equation}
S_{c}[x,\psi,e, \chi] =  \int d\tau
\left ( \frac{1}{2}  e^{-1} (\dot x^\mu -i\chi \psi^\mu)^2  + \frac{i}{2}  \psi_\mu \dot \psi^\mu  \right )\, ,
\end{equation}
whose local symmetries may be recovered from the phase space ones.

Finally, a Wick rotation to Euclidean proper time 
produces the Euclidean action 
\begin{equation}
S_E[x,\psi,e, \chi]  =  \int d\tau
\left ( \frac{1}{2}  e^{-1} (\dot x^\mu- \chi \psi^\mu)^2  + \frac12  \psi_\mu \dot \psi^\mu  \right ).
\end{equation}

The massive case is slightly more subtle. To obtain it we use a method of introducing a mass term starting from the  massless theory formulated in one dimension higher. We denote the extra dimension by $x^5$, and coordinates by  $x^M = (x^\mu, x^5)$, so that indices split as $M=(\mu, 5)$.
The massless spin 1/2 particle in one dimension higher is described by the phase space action 

\begin{equation}
S = \int d\tau 
\left (  
p_M \dot x^M + \frac{i}{2}  \psi_M \dot \psi^M   -  \frac{e}{2}p_{M}p^{M} - i \chi \,p_M \psi^M
\right ).
\label{7.66}
\end{equation}
Now one imposes the constraint\footnote{This constraint Poisson-commutes with the Hamiltonian so does not generate any further constraints.} $p_5= m$, where  $m$ is a constant to be identified as  the  mass of the particle in one dimension lower. The action now takes the form 
\begin{equation}
S = \int d\tau 
\left (  
p_\mu \dot x^\mu + m\dot{x}^{5}+ \frac{i}{2}  \psi_\mu \dot \psi^\mu   + \frac{i}{2}  \psi^5 \dot \psi^5  
- e \frac{1}{2}(p_\mu p^\mu+m^2) - i \chi (p_\mu \psi^\mu+ m \psi^5)
\right ).
\label{7.66v2}
\end{equation}
The term with the coordinate $x^5$ is a total derivative and can be dropped from the action but $\psi^5$ is retained. Let us check that this indeed describes  a free, massive spin 1/2 particle,
at least in even dimensions.
We focus directly on $D=4$ dimensions and note that on top of the operators  in (\ref{7.8a}) one finds 
the extra fermionic operator $\hat\psi^5$ that can be identified
with $\Gamma^5/\sqrt{2}$, where $\Gamma_5$ is the usual chirality matrix obeying $(\Gamma_5)^2 =1$. 
The susy constraint  $p_\mu \psi^\mu+ m \psi^5=0$ becomes at the quantum level
\begin{equation}
( -i \Gamma^\mu \partial_\mu + m \Gamma^5) \Psi =0 \;.
\label{7.67}
\end{equation}
One can multiply this by $ \Gamma^5$  and recognise that the set $\tilde \gamma^\mu = \Gamma^5 \Gamma^\mu$ satisfies the standard (with signature $-+++$) Clifford algebra $\lbrace \tilde{\gamma}^{\mu}, \tilde{\gamma}^{\nu} \rbrace = -2 \eta^{\mu\nu}$ which leads to the massive Dirac equation
\bear
(-i \s \partial + m) \Psi =  (\s p + m) \Psi =0\,,
\label{7diracstandard}
\ear
as required.

However, our goal here is to get the massive Dirac equation through path-integral quantisation. 
Let us start from the action in eq. (\ref{7.66}), suitably Wick rotated to
\begin{equation}
S[x,p,\psi,\psi_5,e,\chi]   = \int d\tau 
\Bigl [
-ip_\mu \dot x^\mu + \frac12  \psi_\mu \dot \psi^\mu  + \frac{1}{2}  \psi^5 \dot \psi^5  +
 \frac{e}{2}(p_\mu p^\mu+m^2) + i \chi (p_\mu \psi^\mu+ m \psi^5)
\Bigr ]\, .
\end{equation}
It enters the path integral as

\begin{equation}
Z \sim  \int   
\frac{{D}x  {D} p  {D} \psi  {D} \psi_5   {D} e {D} \chi
}{\rm Vol(Gauge)}\ \e^{- S[x,p,\psi,\psi_5,e,\chi]}  \;.
\end{equation}
Integrating out the momentum gives the configuration space action. 
Before gauge fixing, and in Euclidean time, it takes the form

\begin{equation}
S[x,\psi, \psi_5, e,\chi] = \int_0^1 d\tau \, \frac12
\left ( e^{-1} (\dot x-\chi \psi)^2 + \psi \dot \psi+  \psi_5 \dot \psi_5+  e m^2 + 2 i \chi m \psi_5
\right )\, ,
\label{8freefermion}
\end{equation}
where we have suppressed obvious indices.
There are two local symmetries to take care of, reparameterisations and local supersymmetry, 
with gauge fields  $e$ and $\chi$, respectively.

We start using the reparameterisation invariance to fix $e(\tau) \equiv 2T$ in that Lagrangian which, on the line, reduces the path integral $\int De(\tau)$ to the proper-time integral with trivial Faddeev-Popov measure $\int_0^{\infty}dT$.  For fixed $T$, we then rescale $\tau \to T\tau $.  The gravitino field $\chi$ is the gauge field for the local worldline supersymmetry, and on an 
interval can be gauge-fixed to a constant Grassmann variable $\Theta$, the super-partner of the global proper-time $T$. 
The gravitino path integral $\int D\chi(\tau)$ then gets replaced by the ordinary Grassmann integral $\int d\Theta$. 

Next, let us consider the terms in the worldline action that depend on the gravitino field $\chi(\tau)$. 
Since $\chi^2(\tau) =0$, those terms can be written as
\bear
S_{\chi} &\equiv & \frac{1}{T}\int_0^Td\tau \chi\Bigl(-\half \dot x\cdot \psi + im\psi_5 \Bigr)\, .
\label{8defSchi}
\ear
We can then use the nilpotency of $\Theta$ to replace the exponential by its argument, and perform the integral:

\bear
\int D\chi \,\e^{-S_{\chi}} = \int d\Theta \,\e^{-S_{\Theta}} = \frac{1}{T}\int_0^Td\tau \Bigl(\half \dot x\cdot \psi - im\psi_5 \Bigr)\, .
\ear
At this stage, we have 

\bear
Z =  \int_0^\infty dT  \,\e^{-m^2 T}
\int_I   {D} x {D} \psi {D} \psi_5  
 \frac{1}{T}\int_0^Td\tau \Bigl(\half \dot x\cdot \psi - im\psi_5 \Bigr)
\,\e^{- \int_0^T d\tau \, 
\left (
 \frac{1}{4}\dot x^2 + \frac{1}{2} \psi \dot \psi+ \frac{1}{2} \psi_5 \dot \psi_5 
\right )}\, .
\nonumber\\
\label{8-ZI}
\ear
We must now think about the boundary conditions to be imposed on the Grassmann fields $\psi(\tau)$ and
$\psi_5(\tau)$. For the coordinate path integral, passing from the closed loop to the open line case
means replacing the homogeneous boundary conditions $x^{\mu}(T) - x^{\mu}(0) = 0$ by inhomogeneous ones, 
\bear
x^{\mu}(T) - x^{\mu}(0) = x'^{\mu} - x^{\mu}\, ,
\label{8inhomo}
\ear
so that we calculate off-diagonal elements of the kernel. Likewise the propagator will depend upon the initial and final spin states, so we should expect that the anti-periodicity condition
\bear
\psi^{\mu}(T) + \psi^{\mu}(0) = 0\, ,
\label{8apbc}
\ear
be replaced by the inhomogeneous (``twisted'') condition
\bear
\psi^{\mu}(T) + \psi^{\mu}(0) = \eta^{\mu}\, ,
\label{8twisted}
\ear
where $\eta^{\mu}$ is a constant external Grassmann vector that should generate the spin structure of the kernel.
But here we run into the following 	subtlety with the variation of the action. The variation of the free particle action is
\begin{align}
\delta S_{\psi} &= \int_0^Td\tau \,\delta \psi_{\mu}\dot\psi^{\mu} 
 + \half \bigl(\psi_{\mu}\delta\psi^{\mu}\bigr)\Big\vert^{\tau = T}_{\tau = 0} .
 \label{8deltaSpsi}
\end{align}
The first term gives us the local equation of motion $\dot\psi^{\mu} = 0$. In the closed loop case, we would have anti-periodic boundary conditions on $\psi$ and $\delta\psi$ which would lead to the vanishing of the surface term in (\ref{8deltaSpsi}).
In the open-line case, instead we have (\ref{8twisted}) but $\delta \psi$ remains anti-periodic,  
resulting in a non-zero contribution from the surface term,
 
 \bear
 \half \bigl(\psi\cdot \delta\psi\bigr)\Big\vert^{\tau = T}_{\tau = 0} = \half\eta \cdot \delta\psi(T) \, .
 \label{8st}
 \ear
If the choice of twisted boundary conditions is to be consistent, 
this non-local term should be cancelled by something. To see what is missing, note that we can
switch from anti-periodic boundary conditions on $\psi(\tau)$ to twisted ones on $\xi(\tau)$ by setting
\bear
\psi^{\mu}(\tau) + \half \eta^{\mu} = \xi^{\mu}(\tau)
\label{psitoxi}
\ear 
and that the result of this transformation can be written as

\bear
\int_0^Td\tau\, \half \psi(\tau)\cdot \dot\psi(\tau) \longrightarrow \int_0^Td\tau\,
\half\xi (\tau) \cdot \dot\xi (\tau)  + \half \xi(T)\cdot \xi(0) \, .
\label{8genbt}
\ear
Under an infinitesimal shift of $\xi(\tau)$, the second term on the right-hand side produces an 
additional term $\half\delta\xi(T)\cdot\xi(0) + \half \xi(T)\cdot\delta\xi (0)$ which is just right to cancel
the surface term in (\ref{8deltaSpsi}) (with $\psi$ now replaced by $\xi$)\footnote{ In the coherent state approach to the spinning particle path integral on the line, there appear similar boundary terms in the action which, unlike in the present case, are local. However, their net effect is the same as we have here; namely, their variation cancel boundary terms coming from the variation of the kinetic action~\cite{deBoer:1995cb, basvan-book}.}. This leads us to understand that the precise version of (\ref{8-ZI}) is 
\begin{align}
Z & =  \int_0^\infty dT  \,\e^{-m^2 T}
\int_{x(0) = x'}^{x(T) = x} \hspace{-15pt} {D} x 
\int_{\xi(T)+\xi(0)=\eta} \hspace{-45pt} {D} \xi \hspace{30pt}
\int_{\xi_5(T)+\xi_5(0)=\eta_5} \hspace{-55pt} {D} \xi_5 \hspace{40pt}
 \frac{1}{T}\int_0^Td\tau \Bigl(\half \dot x\cdot \xi - im\xi_5 \Bigr)
 \nonumber\\
 &\times 
\, \e^{- \int_0^T d\tau \,  \frac{1}{4}\dot x^2}  
\, \e^{- \int_0^T d\tau \, 
\frac{1}{2} \xi \cdot \dot \xi- \frac{1}{2} \xi(T)\cdot \xi(0)}
\, \e^{- \int_0^T d\tau \, 
\frac{1}{2} \xi_5 \dot \xi_5- \frac{1}{2} \xi_5(T)\xi_5(0) }\, .
\label{8ZIfin}
\end{align}

We now turn our attention to the prefactor $(\half \dot x\cdot \xi - im\xi_5)$. In the second term, the
equation of motion $\dot \xi_5 = 0$ means that Ehrenfest's theorem gives 
\bear
\der{}{\tau} \langle \xi_5(\tau) \rangle = 0 \, .
\label{8-dotxi5}
\ear
Thus this term is actually independent of $\tau$, so that we are free to replace it by the average
of its endpoint values, and then apply the boundary conditions:
\bear
\langle \xi_5(\tau) \rangle \longrightarrow \half \bigl(\xi_5(T)+\xi_5(0)\bigr) = \half \eta_5 \, .
\label{8-xi5totheta5}
\ear
Similarly, for the first term we can invoke the above-mentioned fact 
that $Q= \dot x\cdot \xi$ is the conserved charge 
associated to the worldline supersymmetry transformations (\ref{susy}). 
Thus we have also
\bear
\der{}{\tau} \langle \dot x\cdot \xi  \rangle = 0 \, ,
\label{8-dotQ}
\ear
and we use this again to replace the $\tau$ - integrand by the average
of its endpoint values:
\bear
\langle \dot x (\tau) \cdot \xi (\tau) \rangle \longrightarrow \half \bigl(\dot x (T) \cdot \xi (T) +\dot x (0) \cdot \xi (0) \bigr) 
\label{8-Qaverage}
\ear
Now we need to Figure out the effect of a factor $\dot x^{\mu}(T)$ or $\dot x^{\mu}(0)$ inserted into the free 
$x$ - path integral. 
An insertion of $\dot x(T)$ into the free path integral will, after the transformations \eqref{shift},\eqref{defx0}, turn into
\bear
\dot x(T) \longrightarrow \frac{x'- x}{T} + \dot q(T) \, .
\ear
The fluctuation term $\dot q(T)$ leads to an insertion under the path integral over $q$ that is odd in $q$, and thus vanishes. 
Moreover from the explicit result for the free $x$-space propagator, \eqref{bk-like-x} with $N=0$, 
we see that this term could as well be represented as a derivative $-2\partder{}{x'^{\mu}}$, acting on the final point of the
trajectory.
Similarly, an insertion $\dot{x}(0)$ can be represented as a derivative $2\partder{}{x^{\mu}}$ of the amplitude with respect to the initial point $x$ which by translation invariance can be replaced by $-2\partder{}{x'^{\mu}}$. After this, we are ready to use the Grassmann boundary conditions
to replace further 
\bear
\dot x (T) \cdot \xi (T) +\dot x (0) \cdot \xi (0) \longrightarrow -2\partder{}{x'}\cdot \bigl(\xi (T) +  \xi (0)\bigr)
\longrightarrow  -2\partder{}{x'}\cdot \eta \, .
\ear
The prefactor term is now completely expressed in terms of external quantities, and does not involve the path
integral variables any more. Thus the path integrals can now be performed. The Grassmann path integrals 
just yield global normalisation factors, independent even of $\eta$ and $\eta_5$ (as can be seen most simply
by applying the transformation of variables (\ref{psitoxi}) in reverse). 
The $x$ path integral together with the global $T$ integration yields the free scalar propagator $D^{xx'}_0$. Thus we have now simply (up to normalisation)
\bear
Z = \bigl(-i\eta\cdot \partder{}{x} + m\eta_5\bigr) D^{xx'}_0
\ear
The remaining task of matching this to (\ref{StoK}) (for the free case $A=0$) 
parallels our discussion for the operator formalism above.
We require a rule for mapping the Grassmann variables to gamma matrices. 
It would be inconsistent to map $\eta_5$ to $1$ and $\eta^{\mu}$ to $-\gamma^{\mu}$, so we are instead led to identify $\eta_5$ with $\gamma_5$, and reuse the
fact that $\gamma_5\gamma^{\mu}$ are equivalent to $\gamma^{\mu}$ so we finally choose the assignation
\bear
\eta_5\longrightarrow  \gamma_5, \quad \eta^{\mu} \longrightarrow -\gamma_5\gamma^{\mu} \, .
\ear
In this way $Z$ gets mapped into $\gamma_5 S^{xx'}$, rather than $S^{xx'}$, but this is equivalent, and the best we can do. 
It may appear awkward to introduce $\gamma_5$ in this seemingly non-chiral context,
but the fact is that its appearance is a common feature of first-principle approaches to the path integral representation
of the massive fermion  propagator.

\section{Path-ordered path integrals and symbol map}
\label{app-symb}


In this appendix, we prove 
the identity \eqref{idsfgen} 
that allows us to replace the Feynman spin factor \eqref{defspinfactor}
with a path integral over Grassmann variables 
via the symbol map. Our proof essentially follows \cite{fragit}.

First, by standard functional calculus we can rewrite

\bear
\mathcal{P}\left\{e^{-\frac{i}{2}e\int_{0}^{T}d\tau\,\gamma^{\mu}F_{\mu\nu}\gamma^{\nu}}\right\} 
=
\e^{i\frac{e}{2}\int_{0}^{T} d\tau F_{\mu\nu}(x(\tau))\frac{\delta}{\delta
 \theta_{\nu}(\tau)}\frac{\delta}{\delta \theta_{\mu}(\tau)}}
\mathcal{P}
\Bigl\lbrack 
\e^{ \int_{0}^{T} d\tau \, \theta_{\lambda}(\tau)  \gamma^{\lambda}}\Bigr\rbrack
\Big\vert_{\theta = 0} \, \\
\label{idfunctional}
\ear
with Grassmann-valued functions $\theta^{\mu}(\tau)$ that anticommute with the $\gamma^{\mu}$. 

Next, we remove the path-ordering operator using the identity
\bear
\mathcal{P}
\Bigl\lbrack 
\e^{ \int_{0}^{T} d\tau \, \theta(\tau) \cdot \gamma}\Bigr\rbrack
=
\e^{ \int_{0}^{T} d\tau \, \theta(\tau) \cdot \gamma}
\e^{\frac{1}{2}\int_{0}^{T}d\tau \, \int_{0}^{T}d\tau^{\prime}\theta^{\mu}(\tau) 
{\rm sign} (\tau - \tau^{\prime})\delta_{\mu\nu}\theta^{\nu}(\tau^{\prime})}\, .
\label{idremoveP}
\ear
Now on the right-hand side the first exponential can be rewritten as 
\bear
\e^{ \int_{0}^{T} d\tau \, \theta(\tau) \cdot \gamma}
=
\e^{i\frac{\gamma^{\mu}}{\sqrt{2}} \frac{\partial}{\partial \eta^{\mu}}}
\e^{i\sqrt{2}\int_0^T d\tau \, \theta_{\nu}(\tau) \eta^{\nu}}\Bigl\vert_{\eta = 0}\, ,
\ear
where the $\eta^{\mu}, \mu = 1,\ldots, D$ are Grassmann numbers that again must
anticommute with the $\gamma^{\mu}$,
while the second exponential can be replaced by a Gaussian Grassmann path
integral:
\bear
\e^{\frac{1}{2}\int_{0}^{T}d\tau \int_{0}^{T}d\tau^{\prime}\theta^{\mu}(\tau) 
{\rm sign} (\tau - \tau^{\prime})\delta_{\mu\nu}\theta^{\nu}(\tau^{\prime})}
=
\frac
{
\int_{\psi(0) + \psi(T) = 0}{D}\psi \, \e^{-\int_{0}^{T}d\tau \left[ \frac{1}{2}\psi \cdot \dot{\psi} 
	 -i \sqrt{2}\, \theta\cdot \psi\right]  }
}
{
\int_{\psi(0) + \psi(T) = 0}{D}\psi \, \e^{-\int_{0}^{T}d\tau  \frac{1}{2}\psi \cdot \dot{\psi}   }
}\, .
\ear
Here the denominator is the free path-integral normalisation, which is equal to $2^{\frac{D}{2}}$
in $D$ (even) dimensions. Thus the previous three equations can be combined to
\bear
\mathcal{P}
\Bigl\lbrack 
\e^{ \int_{0}^{T} d\tau \, \theta(\tau) \cdot \gamma}\Bigr\rbrack
=
2^{-\frac{D}{2}}
\e^{i\frac{\gamma^{\mu}}{\sqrt{2}} \frac{\partial}{\partial \eta^{\mu}}}
\int_{\psi(0) + \psi(T) = 0}{D}\psi \, \e^{-\int_{0}^{T}d\tau \left[ \frac{1}{2}\psi \cdot \dot{\psi} 
	 -i\sqrt{2} \, \theta_{\mu} (\psi^{\mu}+\eta^{\mu})\right]  }\Bigl\vert_{\eta = 0}\, .
\label{idremovePcombi}
\ear
Now we act on this with the functional operator of \eqref{idfunctional}. This produces

\bear
\e^{i\frac{e}{2}\int_{0}^{T} d\tau F_{\mu\nu}(x(\tau))\frac{\delta}{\delta
 \theta_{\nu}(\tau)}\frac{\delta}{\delta \theta_{\mu}(\tau)}}
\, \e^{-\int_{0}^{T}d\tau \left[ \frac{1}{2}\psi \cdot \dot{\psi} 
	 -i\sqrt{2}\, \theta_{\mu} (\psi^{\mu}+\eta^{\mu})\right]  }
=
\e^{-\int_{0}^{T}d\tau \left[ \frac{1}{2}\psi \cdot \dot{\psi} - i e (\psi^{\mu}+\eta^{\mu}) F_{\mu\nu} 
(\psi^{\nu}+\eta^{\nu})\right]  }\,,
\nonumber\\
\ear
and thus by combining the previous two equations with our starting identity \eqref{idfunctional} we get
\bear
\mathcal{P}\left\{\e^{-\frac{1}{2}e\int_{0}^{T}d\tau\,\gamma^{\mu}F_{\mu\nu}\gamma^{\nu}}\right\} 
=
2^{-\frac{D}{2}}
\e^{i\frac{\gamma^{\mu}}{\sqrt{2}} \frac{\partial}{\partial \eta^{\mu}}}
\int {D}\psi \, \e^{-\int_{0}^{T}d\tau \left[ \frac{1}{2}\psi \cdot \dot{\psi} - i e (\psi^{\mu}+\eta^{\mu})
 F_{\mu\nu} 
(\psi^{\nu}+\eta^{\nu})\right]  }
\Bigl\vert_{\eta = 0}\, .
\nonumber\\
\label{idfinal}
\ear
The final step is to observe that the operation
\bear
\e^{i\frac{\gamma^{\mu}}{\sqrt{2}} \frac{\partial}{\partial \eta^{\mu}}}
f(\eta)
\Bigl\vert_{\eta = 0}
\label{opsymb}
\ear
order by order just corresponds to the replacement of products of $\eta^{\alpha}$s by antisymmetrised
products of $\gamma^{\alpha}$, that is, to the inverse of the symbol map defined in \eqref{defsymb}.
This completes the proof of the identity \eqref{idsfgen}.  

\section{Proof of the hypergeometric identity \eqref{idF}}
\label{app-hypergeo}

In this appendix we show how to reduce the hypergeometric identity \eqref{idF} 
\begin{equation}
	_{2}F_{1}(a, 1; 2-a; -z)(1 - z)(1-2a) +{} _{2}F_{1}(a+1, 2; 2-a; -z)(1 + z)^{2} = 2(1-a)
	\label{idhypergeoapp}
\end{equation}
to known identities.
The arguments of the hypergeometric functions appearing in this identity are of the special kind which makes
it possible to rewrite them in terms of {\it Associated Legendre functions of the first kind} $P^{\mu}_{\nu}(z)$
using the identity (eq. 15.4.15 of \cite{abrasteg-book})
\bear
{}_{2}F_{1}(a,b;a-b+1;z) = \Gamma(a-b+1)(1-z)^{-b}(-z)^{\frac{1}{2}(b-a)}P^{b-a}_{-b}\Bigl(\frac{1+z}{1-z}\Bigr)
\qquad (-\infty < z < 0)\, .
\nonumber\\
\label{15.4.15}
\ear
Applying this identity (with $a$ and $b$ interchanged and $z\to -z$) we find

\begin{eqnarray}
	{}_{2}F_{1}(a, 1; 2-a; -z) &=&  \Gamma(2-a)(1+z)^{-a} z^{\frac{a-1}{2}} P^{a-1}_{-a} \Bigl(\frac{1-z}{1+z}\Bigr)\, ,
\nonumber\\
	{}_{2}F_{1}(a+1, 2; 2-a; -z) &=&  \Gamma(2-a)(1+z)^{-a-1} z^{\frac{a-1}{2}} P^{a-1}_{-a-1} \Bigl(\frac{1-z}{1+z}\Bigr)\, .
	\nonumber\\
	\label{hypergeotolegendre}
\end{eqnarray}
For the Legendre functions one has the ``varying degree identity'' (eq. 8.5.3 of \cite{abrasteg-book})
\bear
(\nu - \mu +1)P^{\mu}_{\nu +1}(x) = (2\nu +1)x P^{\mu}_{\nu}(x) - (\nu + \mu) P_{\nu-1}^{\mu}(x) \, .
\label{8.5.3}
\ear
Using this identity with $\mu=a-1$, $\nu = -a$ and $x = \frac{1-z}{1+z}$ yields
\bear
2(1-a) P^{a-1}_{1-a}\Bigl(\frac{1-z}{1+z}\Bigr)= (1-2a) \frac{1-z}{1+z} P^{a-1}_{-a}\Bigl(\frac{1-z}{1+z}\Bigr) + P_{-a-1}^{a-1} \Bigl(\frac{1-z}{1+z}\Bigr) \, .
\label{idPspecial}
\ear
Multiplying both sides by a factor of $\Gamma(2-a)(1+z)^{1-a} z^{\frac{a-1}{2}}$, and combining the result with \eqref{hypergeotolegendre}, leads to 
\eqref{idhypergeoapp} provided that 
\bear
\Gamma(2-a)(1+z)^{1-a} z^{\frac{a-1}{2}} P^{a-1}_{1-a}\Bigl(\frac{1-z}{1+z}\Bigr) \stackrel{!}{=} 1 \, ,
\ear
which can be verified using the identity (eq. 8.6.16 of \cite{abrasteg-book}),

\bear
P_{\nu}^{-\nu}(x) = \frac{2^{-\nu}(1-x^2)^{\frac{1}{2}\nu}}{\Gamma(\nu +1)}\, ,
\label{8.6.16}
\ear
now with $\nu = 1-a$ (and $x = \frac{1-z}{1+z}$).

\end{document}